# Flexible radiofrequency carbon nanotube transistors operating at frequencies above 100 GHz

Fan Xia[1]†, Tian Xia[1,2]†, Haotian Su[3]†, Lanyue Gan[1], Qianlan Hu[1], Wanyi Wang[1], Ruyi Huang[1,2], Tianshun Bai[1], Yufan Chen[1], Chao Ma[1], Guanhua Long[1], Shan X. Wang[3], Eric Pop[3]*, Lian-Mao Peng[1,2]*, Youfan Hu[1,2]*

[1]Key Laboratory for the Physics and Chemistry of Nanodevices, Center for Carbon-Based Electronics, and School of Electronics, Peking University, Beijing 100871, China.
[2]Academy for Advanced Interdisciplinary Studies, Peking University, Beijing 100871, China.
[3]Department of Electrical Engineering, Stanford University, Stanford, CA 94305, USA.
†These authors contributed equally to this work.
*Corresponding author. Email: epop@stanford.edu; lmpeng@pku.edu.cn; youfanhu@pku.edu.cn

**The development of the sixth generation of wireless communications technology (6G) requires terminals that can operate at frequencies above 100 GHz. For human-centric applications, these terminals should also be flexible and have low power. However, current flexible radiofrequency transistors typically have lower maximum frequencies, in part due to the poor thermal conductivity of flexible substrates. Here, we report radiofrequency transistors that are based on aligned carbon nanotube arrays on flexible substrates and have current gain cutoff frequencies ($f_T$) and power gain cutoff frequencies ($f_{max}$) above 100 GHz. This is achieved by using electro-thermal co-design to improve the heat dissipation and radiofrequency performance of the devices. The transistors exhibit an on-state current of 0.947 mA $\mu m^{-1}$, a transconductance of 0.728 mS $\mu m^{-1}$, a peak extrinsic $f_T$ of 152 GHz, a peak extrinsic $f_{max}$ of 102 GHz, and a power consumption under 200 mW $mm^{-1}$. We also show that the devices can be used to create flexible radiofrequency amplifiers with an output power of 64 mW $mm^{-1}$ and a 11 dB power gain in the K-band.**

The sixth generation of wireless communications technology (6G), which is expected to have data transmission rates in excess of 100 gigabits per second, will require devices with wide bandwidths and that can operate at frequencies within the terahertz (THz) range of 0.1 THz to 10 THz[1-5]. A particular focus of 6G technology is on highly connected, human-centric communications[5-7], such as remote terminals that can monitor physiological signals or replicate human senses, and then share them[8,9]. This requires the technology to also be light weight, biocompatible, and adaptable[10].

Flexible wireless terminals that operate at frequencies above 100 gigahertz (GHz) with low power consumption are ideal for human-centric 6G applications, and flexible radiofrequency (RF) transistors are a key component of such systems. Flexible RF transistors based on III-V materials have demonstrated operating frequencies exceeding 30 GHz[11]. However, their incompatibility with complementary metal-oxide-semiconductor (CMOS) technology poses a challenge for integration. Among previously reported flexible RF metal-oxide-semiconductor field-effect transistors (MOSFETs) based on organic materials[12], amorphous oxide semiconductors[13,14] and two-dimensional (2D) materials[15-18], those based on graphene typically offer the highest performance, with reported extrinsic current-gain cutoff frequencies ($f_T$) of up to 56 GHz[15]. However, making flexible RF MOSFETs capable of operating beyond 100 GHz with low power remains a challenge.

Carbon nanotubes (CNTs) are promising candidates for next-generation integrated circuits[19,20], with theoretical estimates suggesting that CNT-based MOSFETs could work in the THz regime due to their high saturation velocity, small intrinsic capacitance, and high thermal conductivity and stability[21-23]. CNT-based RF transistors have been created on rigid substrates[24-27] with an extrinsic $f_T$ of up to 186 GHz[27]. In addition, CNT-based RF transistors can be easily integrated with digital and analogue circuits based on the same material, which is critical for system-on-chip technology[28]. The moderate bandgaps of nanotubes also enable low-power operation[20], facilitating integration with biointerfaces. However, the highest reported $f_T$ of flexible CNT-based RF transistors is currently only 1.6 GHz[29], far below their theoretical potential.

In this Article, we report high-performance flexible RF transistors that are based on aligned nanotube arrays and designed with electro-thermal co-optimization on flexible polyimide (PI) substrates. The transistors exhibit extrinsic $f_T$ and power-gain cutoff frequency ($f_{max}$) values of 152 GHz and 102 GHz, respectively. They also have low power consumption, with direct current (d.c.) power of 199 mW $mm^{-1}$ and 147 mW $mm^{-1}$ (for the highest $f_T$ and $f_{max}$, respectively). We use these transistors to create flexible RF power amplifiers operating at 18 GHz, with a gain of 11 dB and output power of 64.45 mW $mm^{-1}$.

**Device Structure and d.c. Performance**

Human-centric communication in the 6G technology envisions multiple seamlessly integrable terminals bonded to the human body for information collection, sharing, and communication, as illustrated in Fig. 1a. The structure of our RF transistors based on aligned CNTs arrays is depicted in Fig. 1b, where the transistors are fabricated on a flexible polyimide (PI) substrate with a thickness of 2 μm. The design incorporates a two-finger, top-gated architecture with air gaps between the gate and source/drain contacts to reduce parasitic effects. Owing to its low dielectric loss, high glass temperature, relatively high thermal conductivity (among flexible polymer substrates)[30,31], the PI substrate is selected to meet the stringent requirements of RF applications. Its low surface roughness (average roughness 0.27 nm, Supplementary Fig. 1), comparable to that of typical silicon substrates, was achieved via spin-coating and facilitates the fabrication of high-performance devices. Aligned CNT arrays were deposited on a silicon wafer via a dimension-limited self-alignment procedure[27,32] and then transferred to the PI substrate. Palladium (Pd) was chosen as the contact metal to form ohmic contacts with the CNTs[32,33]. The entire fabrication process is provided in Supplementary Fig. 2.

Scaling down the channel length ($L_{ch}$) is an effective way to promote transistor performance[33,34]. Moreover, achieving impedance matching between the transistor and the measurement system requires a large channel width, while minimizing parasitic capacitance necessitates air gaps between the gate and the source/drain contacts. To meet these requirements, the gate electrode must possess a high width–to–length ratio and be precisely aligned at the center of the downscaled channel. To address these challenges, we optimized the fabrication process on the flexible substrate, achieving a gate length ($L_g$) down to 75 nm. Each finger of the two-finger device features a channel width of 10 μm, resulting in a width-to-length ratio of 133:1. Figure 1c shows scanning electron microscopy (SEM) images of the fabricated flexible RF transistors. In the enlarged view at the bottom, the 75 nm-long gate is well-aligned at the center of the 120 nm-long channel. The SEM image of the aligned CNT arrays on the flexible substrate is shown in Supplementary Fig. 3, demonstrating a high density of approximately 150 CNTs per μm and a good uniformity.

The transfer and output characteristics of a typical flexible RF transistor with an $L_g$ of 75 nm and an $L_{ch}$ of 120 nm are presented in Fig. 1d,e. The device exhibits excellent d.c. performance, achieving an on-state current ($I_{on}$) of 0.947 mA/μm and a transconductance ($g_m$) of 0.728 mS/μm. These values are comparable to those achieved in aligned CNT array-based RF transistors on rigid substrates under similar bias conditions[27]. The intrinsic $f_T$ ($f_{T,\,int}$) is theoretically proportional to $g_m$[21,35,36], indicating the importance of a large $g_m$ in d.c. performance for achieving superior RF performance in flexible RF transistors. Compared with other state-of-the-art high-performance flexible MOSFETs based on various materials[14,16,17,34,37-39], as shown in Fig. 1f and Supplementary Table 1, our devices exhibit the highest $I_{on}$ and $g_m$ at supply voltage no more than 0.6 V. This highlights their potential for enabling high-performance, low-power flexible electronics.

**Electro-Thermal Co-Design**

Achieving very high current densities is often challenging for downscaled transistors due to significant self-heating, which presents a critical bottleneck to performance and reliability[40-42]. This issue is particularly pronounced in flexible devices with polymer-based substrates, which have low thermal conductivity[37,43]. The breakdown temperature of CNTs in air[41,44] is approximately 600 °C, while the glass transition temperature of the PI substrate is approximately 350 °C, with a decomposition temperature[31] of nearly 500 °C. To assess the thermal constraints imposed by the flexible substrate, we conducted finite-element electro-thermal simulations[37] to compare the temperature profiles of an electrically optimized CNT-based RF transistor[27] operating at the same power on a rigid quartz substrate and a flexible PI substrate. As shown in Extended Data Fig. 1, the transistor operates well below its breakdown temperature on the rigid substrate. However, on the PI substrate, the maximum temperatures of the CNTs and the PI substrate reached 627 °C and 341 °C, respectively, exceeding or approaching their breakdown limits. These findings highlight that, on flexible substrates, the high performance of RF devices requires not only the superior material properties of the aligned CNT arrays, but it is also essential to limit their self-heating by reducing power consumption and implementing effective thermal management strategies.

We employed electro-thermal co-design for flexible RF transistors to achieve high RF performance and low power consumption, while maintaining a well-controlled temperature rise ($\Delta T$). For transistors on flexible substrates with low thermal conductivity, heat dissipation through the contacts becomes critical[37]. Thicker metal contacts aid heat dissipation both due to higher thermal conductance (larger area for heat flow) and higher thermal conductivity (less boundary scattering)[45], while thinner contacts minimize parasitic capacitances between source/drain and gate, for high-speed performance. To balance these requirements, a contact stack of 20 nm palladium (Pd) and 20 nm gold (Au) was employed (Fig. 2a), where Pd forms good ohmic contact with CNTs, and Au improves thermal and electrical conductance. The gate was similarly engineered using an aluminum oxide ($Al_2O_3$)/aluminum (Al)/titanium (Ti)/Au stack to minimize thermal resistance and provide an additional heat dissipation pathway. $Al_2O_3$ was selected as the gate dielectric for its higher thermal conductivity than hafnium oxide ($HfO_2$)[46], with an Al layer directly deposited on top to achieve a thermal boundary conductance (TBC)[47] of ~200 $MWm^{-2}K^{-1}$, significantly better than expected for Au/$Al_2O_3$ interfaces with a TBC of ~20 $MWm^{-2}K^{-1}$. A thin Ti layer was introduced between Al and Au to prevent the formation of Au-Al intermetallic, which may introduce thermal and electrical instability[48]. To further optimize heat dissipation, the total gate stack thickness was set to 190 nm, ensuring near-bulk metal thermal conductivity[45] while simultaneously reducing gate resistance for improved RF performance. These design strategies are summarized in Extended Data Fig. 2, and their quantitative electro-thermal trade-offs are further analyzed in Supplementary Note 1 and Supplementary Table 2.

The heat map in Fig. 2a illustrates the calculated $\Delta T$ of the flexible RF transistor with $L_g$ = 75 nm under working conditions, showing a maximum $\Delta T$ of 331 °C for the CNT channel and 168 °C for the flexible substrate. Thermal analysis details are provided in Extended Data Fig. 3 and Supplementary Note 2. (the validity of these simulated temperature rises is experimentally anchored by CNT breakdown measurements, with the results shown in Supplementary Fig. 4 and a detailed analysis provided in Supplementary Note 3). In the zoomed-in, top-down view of the channel (Fig. 2b, top), the temperature profile appears uniform across its width direction, with significant temperature drops at the source and drain contacts. As shown in the $\Delta T$ profile along the channel (Fig. 2b, bottom), the temperature decays with a characteristic thermal healing length ($L_H \approx$ 300 nm), which is much longer than that on rigid substrates with better heat-sinking properties (e.g., doped Si). This underscores the critical role of source and drain contacts for lateral heat dissipation, suggesting that the CNT contact[49] length should ideally exceed $L_H$.

We evaluated the maximum $\Delta T$ of flexible RF transistors with $L_g$ ranging from 75 nm to 420 nm, as summarized in Fig. 2c. Across all devices, the $\Delta T$ of the CNTs and the flexible substrate remain below 360 °C and 200 °C, respectively. The maximum $\Delta T$ shows slight variations corresponding to changes in d.c. power ($P_{dc} = I_d V_{ds}$ during the scattering (S) parameter measurements), as shown in Extended Data Fig. 3c. All $L_{ch}$ here are relatively short, $L_{ch} < 3L_H$ (with $L_{ch}$ ranging from 120 to 540 nm), which explains why the thermal resistance of our devices does not scale with $L_{ch}$[49]. This is reflected in Fig. 2c where the peak temperature is relatively constant (at relatively constant power) despite the channel length increasing by > 4×. For longer channel devices ($L_{ch} > 3L_H$) or devices on silicon (where $L_H$ is much shorter), one would expect the peak temperature to scale inversely with the channel length, given the same power, because the power density (per channel area) would reduce. These results confirm that, with our optimized device design for heat dissipation, the CNT channel operates well below its breakdown temperature, while the maximum temperature of the PI substrate remains safely below its glass transition temperature. In principle, our devices could sustain even higher $P_{dc}$, offering the potential for higher RF performance without thermal degradation.

**RF Performance**

Figure 2d shows the relationship between the current gain and frequency for a flexible transistor with a $L_g$ of 75 nm under small-signal measurements at an input power of -10 dBm, extracted from the measured two-port S-parameters shown in Extended Data Fig. 4. The extrinsic (as-measured) $f_T$ ($f_{T,ext}$) of the device reaches 152 GHz, falling within the sub-THz regime (100 GHz to 300 GHz). After performing a standard pad de-embedding procedure (more details in Supplementary Note 4 and Extended Data Fig. 5), the pad de-embedding $f_T$ ($f_{T,pad}$) reaches 171 GHz, exceeding that of 90 nm node RF-CMOS devices ($L_g$ = 65 nm, $f_{T,pad}$ = 160 GHz) [50], with a low de-embedding ratio ($f_{T,pad}/f_{T,ext}$) of 1.125, indicating minimal parasitic effects from the substrate and the measurement setup. The intrinsic de-embedding further eliminates the parasitic effects from the device structure itself, revealing the intrinsic material properties[17,27]. This yields an intrinsic $f_T$ ($f_{T,int}$) of 310 GHz, which enters the THz regime (300 GHz to 3 THz), and is comparable to the $f_{T,int}$ obtained from RF transistors based on aligned CNT arrays on silicon wafer with similar gate lengths ($L_g$ = 60 nm, $f_{T,int}$ = 365 GHz; $L_g$ = 90 nm, $f_{T,int}$ = 238 GHz) [27]. This demonstrates that the aligned CNT arrays retain their superior properties after being transferred from the silicon wafer to the PI substrate. Details of the extrapolation method and simulation results are provided in Supplementary Fig. 5, Extended Data Fig. 6, and Supplementary Tables 3,4.

Figure 2e presents the small-signal power gains as a function of frequency for a flexible transistor with $L_g$ = 160 nm. The extrinsic $f_{max}$ ($f_{max,ext}$) maximizes in this device, reaching 102 GHz. After pad de-embedding and intrinsic de-embedding, the $f_{max}$ reaches 141 GHz ($f_{max,pad}$) and 168 GHz ($f_{max,int}$), respectively. Figure 2f shows $f_{T,ext}$ under varying bias conditions ($V_{gs}$ and $V_{ds}$). A higher $V_{ds}$ offers higher $f_{T,ext}$, while the optimal $V_{gs}$ (approximately -0.5 V) corresponds to the $V_{gs}$ that maximum $g_m$ in the d.c. performance (Fig. 1d). The optimized bias conditions for achieving the highest $f_{max,ext}$, $f_{T,pad}$, $f_{max,pad}$ are summarized in Extended Data Fig. 7.

To highlight the impact of our electro-thermal co-design, Supplementary Fig. 6 presents the output characteristics of a flexible RF transistor based on aligned CNTs *without* such co-design. Pronounced negative differential resistance is observed under high-bias conditions due to severe self-heating, which significantly degrades performance and reliability. These devices only achieved a maximum $f_{T,ext}$ of 44 GHz ($L_g$ = 75 nm) and $f_{max,ext}$ of 30 GHz ($L_g$ = 100 nm), as shown in Extended Data Fig. 8, substantially lower than those *with* electro-thermal co-design. Furthermore, over half of these devices experienced immediate breakdown after RF measurements, as shown in Supplementary Fig. 7. In contrast, flexible RF transistors with electro-thermal co-design consistently delivered higher performance and exhibited excellent reliability and endurance across repeated RF and d.c. measurements (Supplementary Fig. 8), as well as good shelf-life stability (Supplementary Fig. 9).

To investigate the scaling behavior, the extrinsic current gains and power gains of flexible RF transistors with six different $L_g$ ranging from 75 nm to 420 nm are shown in Fig. 3a, b, respectively, while the pad de-embedded and intrinsic current gains and power gains are presented in Extended Data Fig. 9. Figure 3c illustrates the relationship between $f_T$ (extrinsic, pad de-embedded, and intrinsic) and $L_g$. Both the extrinsic and pad de-embedded $f_T$ increase as $L_g$ decreases, while the intrinsic $f_T$ shows inversely proportionality to $L_g$, consistent with the small signal model ($f_{T,int} = g_m/(2\pi C_g) = g_m/(2\pi C_{ox} W L_g) \propto 1/L_g$, where $C_g$ is the gate capacitance, $C_{ox}$ is the gate capacitance per unit area and $W$ is the channel width) [51]. Figure 3d depicts the variations in $f_{max}$ (extrinsic, pad de-embedded, and intrinsic) with $L_g$. Notably, $f_{max}$ peaks at $L_g$ = 160 nm and decreases as $L_g$ further decreases, which can be attributed to the rapid increase in gate resistance ($R_g$) in shorter-channel devices. The extraction of $R_g$ and a detailed discussion of its impact on $f_{max}$ performance are provided in Supplementary Note 5 and Supplementary Fig. 10.

Figure 3e shows the extracted de-embedding ratios from Fig. 3c,d, which are defined as the ratio of the de-embedded to the corresponding extrinsic $f_T$ or $f_{max}$. The pad de-embedding ratios remain below 1.2 for $f_T$ and 1.4 for $f_{max}$ during $L_g$ scaling, which are relatively low compared with those reported in the literature, as summarized in Supplementary Table 5. We measured the two-port S-parameters of the pad de-embedding structures on the PI substrate, with the results shown as Smith charts in Supplementary Fig. 11. The $S_{11}$ and $S_{22}$ parameters of the open structure are very close to the infinite impedance point with negligible frequency dispersion. This finding indicates that the high frequency loss from the PI substrate is very low, benefiting from both the intrinsic dielectric properties of PI and the optimized baking protocol described in the Methods, supporting the low pad de-embedding ratios achieved in these devices. The intrinsic de-embedding ratio of $f_T$ increases as $L_g$ decreases, with more pronounced changes when $L_g$ scales down to sub-200 nm. This indicates that the parasitic effects from the device structure itself become dominant at smaller dimensions, requiring further structure optimization to fully leverage the superior material properties for higher $f_T$. Figure 3f shows the statistical distributions of the $f_{T,ext}$ for flexible RF transistors with different $L_g$, while the $f_{max,ext}$, $f_{T,pad}$ and $f_{max,pad}$ distributions are provided in

Extended Data Fig. 10. The average values and standard variations of $f_T$ and $f_{max}$ are summarized in Supplementary Table 6. Specifically, the $f_{T,ext}$ distribution for flexible RF transistors with $L_g$ = 75 nm is 141.17 ± 9.2 GHz, while the $f_{max,ext}$ distribution of flexible RF transistors with $L_g$ = 160 nm is 92.34 ± 9.23 GHz, demonstrating good uniformity and high performance across the fabricated devices. The device yield and process reliability across different batches are presented in Supplementary Fig. 12, showing consistent performance trends and yields that, although reduced at shorter $L_g$, remain above 60%.

**Performance Benchmarking**

The performance and power consumption of our flexible RF transistors are benchmarked with other flexible RF MOSFETs and relevant CNT-based RF transistors on rigid substrates in Fig. 4 and Supplementary Table 5. As shown in Fig. 4a, our devices achieve the highest $f_{T,ext}$ of 152 GHz compared to other state-of-the-art flexible RF MOSFETs based on various materials, including amorphous oxide semiconductors[13,14], 2D materials[15-18], organic materials[12] and other CNTs[29]. Remarkably, this $f_{T,ext}$ is ~100× higher than that of previously reported flexible RF transistors based on CNT materials[29] and ~3× that of flexible RF transistors based on graphene[15]. Furthermore, our devices achieve the highest $f_{max,ext}$ of 102 GHz, exceeding that of flexible RF MOSFETs based on all other materials[12-18] by more than 7×. These results, for the first time, demonstrate both $f_{T,ext}$ and $f_{max,ext}$ exceeding 100 GHz in flexible RF MOSFETs, a significant milestone for 6G applications. Figure 4b shows the $f_{T,ext}$ versus $f_{max,ext}$ for these benchmarked works, with the geometric mean value $(f_{T,ext} \times f_{max,ext})^{1/2}$ (a balanced indicator of high frequency performance of RF transistors) of our devices reaching 106 GHz, which is approximately 4 times the highest value of flexible RF transistors based on graphene[15].

Low power consumption is another critical requirement for 6G communication terminals. Compared to other flexible RF transistors that typically operate at a supply voltage above 1.5 V[12,13,17,18,29], our devices achieve an $f_{T,ext}$ of 152 GHz under a $V_{ds}$ of 0.65 V and an $f_{max,ext}$ of 102 GHz under a $V_{ds}$ of 1.05 V, as shown in Fig. 4c, showcasing significantly higher performance with lower power consumption. As illustrated in Fig. 4d, the $f_{T,ext}$ and $f_{max,ext}$ versus the $P_{dc}$ are also benchmarked, with $P_{dc}$ values of 199 mW/mm and 147 mW/mm for the highest $f_{T,ext}$ and $f_{max,ext}$, respectively. These values outperform those of other flexible RF MOSFETs based on various materials[12,14,15,17,29] and are even lower than those of their rigid counterparts based on aligned CNTs with similar performance[27,52,53]. These results highlight the potential of our devices for energy-efficient, high-performance flexible RF applications.

**Flexibility Assessment and RF Amplifier Performance**

Flexibility tests were performed by bending the flexible substrate to various curvatures, followed by device characterization after release. Details of the testing setup are provided in the Methods and Supplementary Fig. 13. Figure 5a shows the current gains of a typical RF transistor with $L_g$ = 75 nm after bending to different radii of curvature, along with the changes in $f_{T,ext}$ versus the bending radius. The $f_{T,ext}$ was 142 GHz before bending and degraded by 6.4% after bending to a curvature radius of 1.5 mm. Figure 5b shows the measured current gains and the changes in $f_{T,ext}$ of the RF transistor after multiple bending cycles at a curvature radius of 3 mm. After 1000 bending cycles, $f_{T,ext}$ decreases slightly by 9.8%, demonstrating good mechanical resilience. Under practical deformed conditions, accumulated mechanical stress may degrade long-term reliability through mechano–electrical coupling. This can be mitigated by integrating mechanical and electro-thermal co-design, including positioning devices near the neutral mechanical plane, optimizing device layout to minimize strain concentration, and employing strain-tolerant interconnects and encapsulation.

A flexible RF power amplifier was constructed based on these RF transistors with $L_g$ = 160 nm, incorporating 6 fingers to increase the channel width and achieve impedance match. Figure 5c represents the output power ($P_{out}$) and the power gain of the flexible RF power amplifier versus the input power ($P_{in}$) at a working frequency ($f_0$) of 18 GHz (in the K band). The amplifier achieves a $P_{out}$ of 64 mW/mm and a maximum power gain of 11 dB. Figure 5d shows the drain efficiency (DE) and the power-added efficiency (PAE) of the flexible RF power amplifier versus $P_{in}$, with DE peaking at 18% and PAE at 10%. Our flexible RF power amplifier has good performance when it works under a $f_0$ of 18 GHz, which shows the potential to build flexible RF circuits and communication systems with high frequencies. The pad de-embedded small-signal power gain as a function of frequency for the flexible power amplifier is shown in Supplementary Fig. 14, yielding an $f_{max,pad}$ of 131 GHz. This suggests a potential working frequency of up to 43.6 GHz (one-third of $f_{max,pad}$[54]), at which the corresponding power gain is 8.9 dB.

## Conclusions

We have reported flexible RF transistors that are based on aligned nanotube arrays and optimized through electro-thermal co-design to achieve high RF performance and enhanced heat dissipation. Our flexible transistors exhibited an $I_{on}$ of 0.947 mA μm$^{-1}$ and a $g_m$ of 0.728 mS μm$^{-1}$, demonstrating an extrinsic $f_T$ of 152 GHz and $f_{max}$ of 102 GHz. The devices also showed power consumption below 200 mW mm$^{-1}$. We used the transistors to create a flexible RF power amplifier with a $P_{out}$ of 64 mW mm$^{-1}$ and power gain of 11 dB at a $f_0$ of 18 GHz. Our results show that nanotubes are an effective material for making flexible RF devices on flexible substrates with extrinsic frequencies above 100 GHz.

Further advances will rely on substrate engineering, electro–thermal–mechanical co-design, and device optimization. In particular, it is critical to improve thermal and mechanical robustness for mitigating self-heating and mechano–electrical coupling, thereby providing a stable foundation for RF scaling. On this basis, channel length reduction and advanced gate engineering can be more effectively translated into gains in $f_T$ and $f_{max}$, while improvements in nanotube array quality and density will also be required (Supplementary Note 6 and Supplementary Fig. 15). Eventually, flexible nanotube-based RF circuits could be integrated with nanotube-based flexible sensors, antennas and digital/analogue circuits to achieve high speed and low power terminals for next-generation 6G technology.

## Methods

### Fabrication of the flexible substrates

First, the PI solution (593052, Sigma-Aldrich) was spin-coated onto a rigid temporary supporting substrate (silicon wafer) at a speed of 5000 rpm. Then, a step baking process was carried out to remove the solvent and complete the polyimide imidization process. The optimized baking protocol included three steps: 100 °C for 10 min, 250 °C for 60 min, and 375 °C for 60 min. Finally, a 10-nm-thick $HfO_2$ layer was deposited on the cured PI surface using atomic layer deposition (ALD) (TFS 200, Beneq Oy.) at 90 °C. This $HfO_2$ layer enhanced the adhesive strength between the metal layer and the flexible substrate and improved heat dissipation.

### Transfer of the aligned CNT arrays to the flexible substrate

Poly(methyl methacrylate) (PMMA) was spin-coated onto the aligned CNT arrays on the silicon wafer. The sample was then immersed in hydrofluoric acid for 7 days to etch the surface $SiO_2$ layer and detach the aligned CNT arrays from the wafer. Once separated, the CNT film was cleaned by deionized water and transferred onto the target flexible substrate. The sample was then baked at 60 °C for 10 min to stabilize the transfer, followed by soaking in acetone for 2 h to dissolve the PMMA film.

Fabrication of flexible RF transistors and power amplifiers

First, Ti/Au bilayer films (10/80 nm) were patterned to form alignment markers and interconnects by electron beam lithography (EBL) (Voyager, Raith GmbH), electron beam evaporation (EBE) (DE 400, DE Technology), and a standard lift-off process. Reactive ion etching (Minilock, Trion) was then performed to etch away the unwanted CNTs outside the EBL-defined channel regions. Pd/Au films (20/20 nm) were then deposited to form the source and drain electrodes. A 2.5 nm Al layer was subsequently deposited and oxidized in air to form $Al_2O_3$ as the gate dielectric. Al/Ti/Au films (20/5/165 nm) were deposited to form the gate electrodes. For interconnects and RF input and output pads, Ti/Au films (20/400 nm) were deposited. For power amplifiers with 6 fingers, overexposed PMMA (50k, 12%) bridges were utilized as jumpers between the lower and upper interconnects. Finally, the complete flexible electronic chip was removed from the temporary supporting substrate using a capillary-assisted electrochemical delamination method, as described below.

Capillary-assisted electrochemical delamination

A NaCl electrolyte solution (1 mol $L^{-1}$) was first prepared. After trimming off the fringes of the PI layer to expose the surface of the silicon wafer, the sample was fixed at 45° by clamping equipment, with the electronic devices facing upward. The lower edge of the silicon wafer was then inserted in the NaCl solution. A positive potential of 20 V was applied to the silicon wafer, while the NaCl solution was grounded at 0V. This setup initiated the anodic reaction of $Si - 8e^{-} + 8OH^{-} \rightarrow H_2SiO_3\downarrow + 3H_2O + 2O_2\uparrow$, generating gaps between the PI layer and silicon wafer. The resulting capillary force allowed the NaCl solution to climb upwards, facilitating the delamination of the PI along the inclined silicon wafer.

Measurement of the flexible RF transistors and power amplifiers

The flexible electronic chip was placed flat on a supporting quartz substrate for probe testing. The setup for S-parameter measurements included a probe station (Cascade Summit 200), ground-signal-ground (GSG) probes, coaxial cables, a semiconductor analyzer (Agilent B1500), and a vector network analyzer (Keysight N5247B). The GSG probes and coaxial cables were first calibrated using the standard off-wafer short–open–load–through (SOLT) procedure. The semiconductor analyzer provided the d.c. bias, while the vector network analyzer supplied the RF signal. For power amplifier measurements, a pair of impedance tuners (FOCUS L4030) was used for impedance matching.

Flexibility tests of the flexible RF transistors

The flexible electronic chip was laminated onto acrylic cylinders with different radii for the static bending test. The bending radius was defined by the cylinder radius, which was measured using a micrometer. For cyclic bending tests, the chip was mounted on a supporting PI film, each end of the film was fixed to acrylic plates attached to a translation stage. Periodic motion of the stage then induced repeated bending and release of the flexible electronic chip. The corresponding bending radius during cycling was determined using a graduated ruler.

**Data availability**

The data supporting the findings of this study are available within the article and the Supplementary Information. Source data are provided with this paper, which have been deposited in Figshare at https://doi.org/10.6084/m9.figshare.31709908.

**Acknowledgements**

This work was supported by the National Key R&D Program of China (Grant No. 2022YFB4401603 and 2021YFA1202904 for Y.H.), China Postdoctoral Science Foundation (Grant No. 2025T180148 for F.X.) and Peking Nanofab. H.S. and E.P. acknowledge partial support from SUPREME, one of the seven JUMP 2.0 centers sponsored by Semiconductor Research Corporation (SRC) and DARPA.

**Author contributions**

Y.H. and L.-M.P. proposed and supervised the project. F.X. and T.X designed flexible RF devices and conducted device fabrication. H.S. performed the finite-element electro-thermal modeling. F.X., T.X., Q.H., T.B., and G.L. performed the S-parameter measurements. L.G., W.W. and C.M. performed the flexibility tests. L.G., R.H., and Y.C. prepared flexible substrates. F.X., T.X, H.S., E.P., L.-M.P., and Y.H. analyzed the data and co-wrote the manuscript. All the authors discussed the results and revised the manuscript.

**Competing interests**

The authors declare no competing interests.

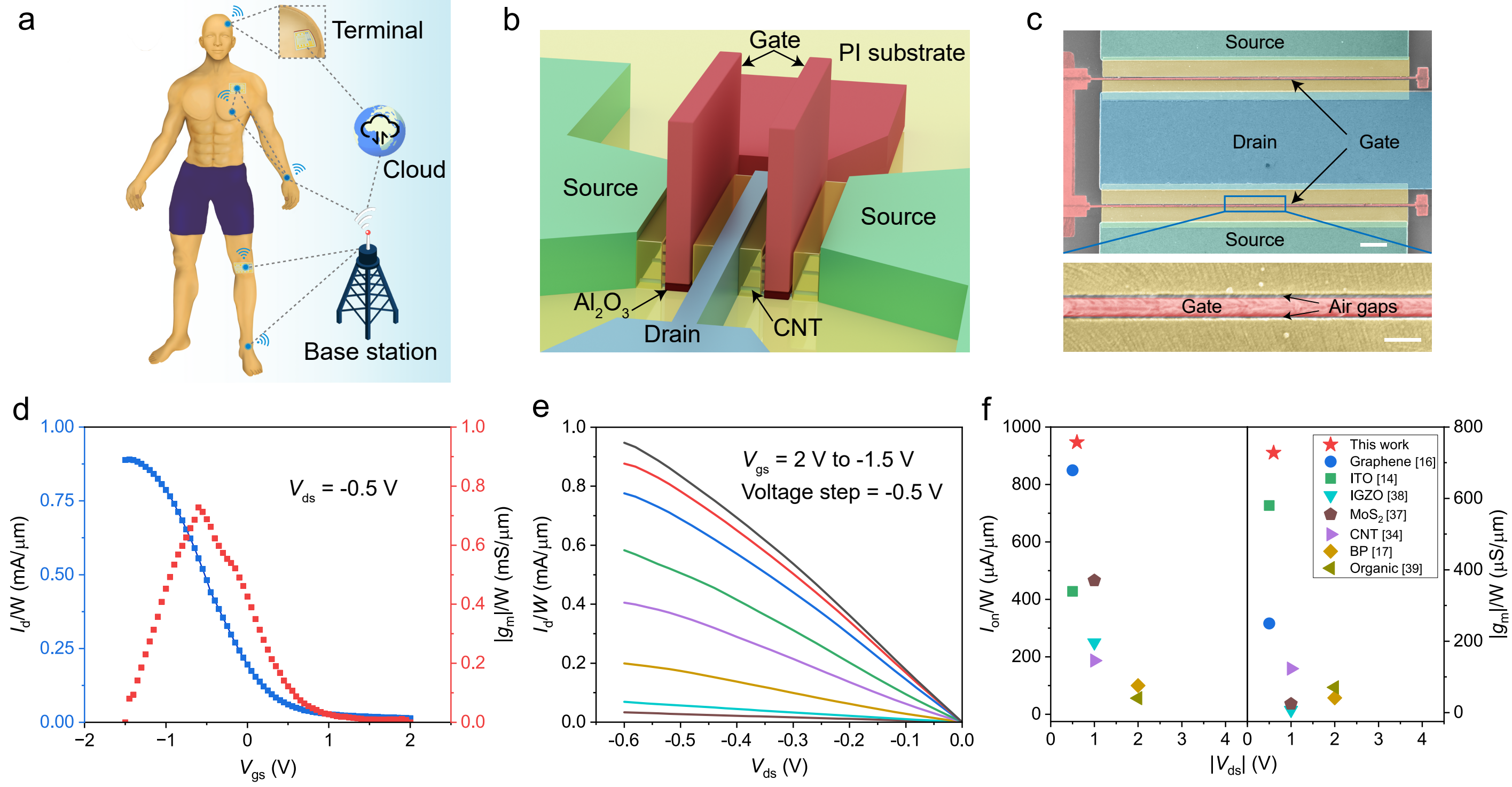

**Fig. 1 | Structure and d.c. characteristics of flexible CNT-based RF transistors. a**, Conceptual illustration of seamlessly integrable terminals for human-centric communications in 6G technology. **b**, Schematic of the structure of a flexible RF transistor based on aligned CNT arrays (not drawn to scale). **c**, Top: SEM image of a two-fingered flexible RF transistor with $L_g$ = 75 nm and $L_{ch}$ = 120 nm. Scale bar: 1 μm. Bottom: Zoomed-in SEM image of the channel area. Scale bar: 300 nm. **d**, Transfer characteristics (blue curve) and transconductance (red curve) for a flexible RF transistor with $L_g$ = 75 nm and $L_{ch}$ = 120 nm. **e**, Output characteristics of the same transistor in **c**. **f**, Comparison of $I_{on}$ and $g_m$ versus the absolute value of $V_{ds}$ between our devices and other state-of-the-art flexible RF transistors based on various materials, including indium tin oxide (ITO) [14], indium gallium zinc oxide (IGZO) [38], molybdenum disulfide ($MoS_2$) [37], CNT[34], black phosphorus (BP) [17], graphene[16], and organic materials[39].

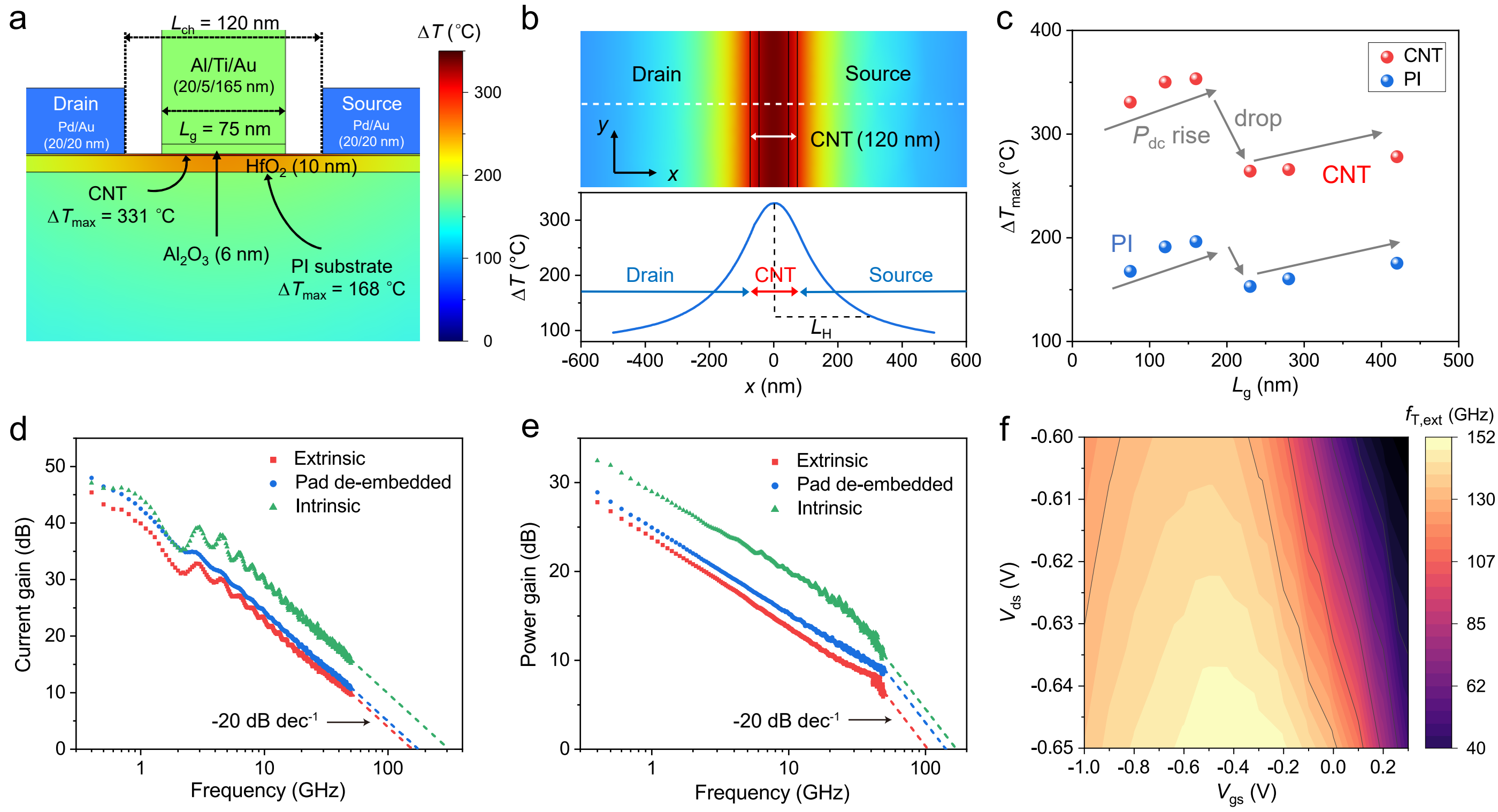

**Fig. 2 | Electro-thermal co-design and performance of flexible CNT-based RF transistors. a**, Heat map showing the temperature rise ($\Delta T$) of a flexible RF transistor with $L_g$ = 75 nm at a power of 1.99 mW. The power was measured under bias conditions when the maximum $f_T$ was reached. **b**, Top: Zoomed-in *x*-*y* cross-section view of the device in **a** around the channel area. Bottom: $\Delta T$ profile along the CNT channel direction (dashed line in the top figure), showing the temperature decay over a characteristic thermal length $L_H \approx 300$ nm. **c**, Maximum $\Delta T$ of the CNT channel and the PI substrate with different $L_g$. The input power varies here for different $L_g$, as shown in Extended Data Fig. 3c. **d**, Extrinsic, pad de-embedded, and intrinsic current gains versus frequency for the champion flexible RF transistor with $L_g$ = 75 nm. **e**, Extrinsic, pad de-embedded, and intrinsic power gains versus frequency for the champion flexible RF transistor with $L_g$ = 160 nm. The slope of the extrapolation line is -20 dB dec$^{-1}$. **f**, $f_{T,ext}$ as a function of bias conditions, $V_{gs}$ and $V_{ds}$.

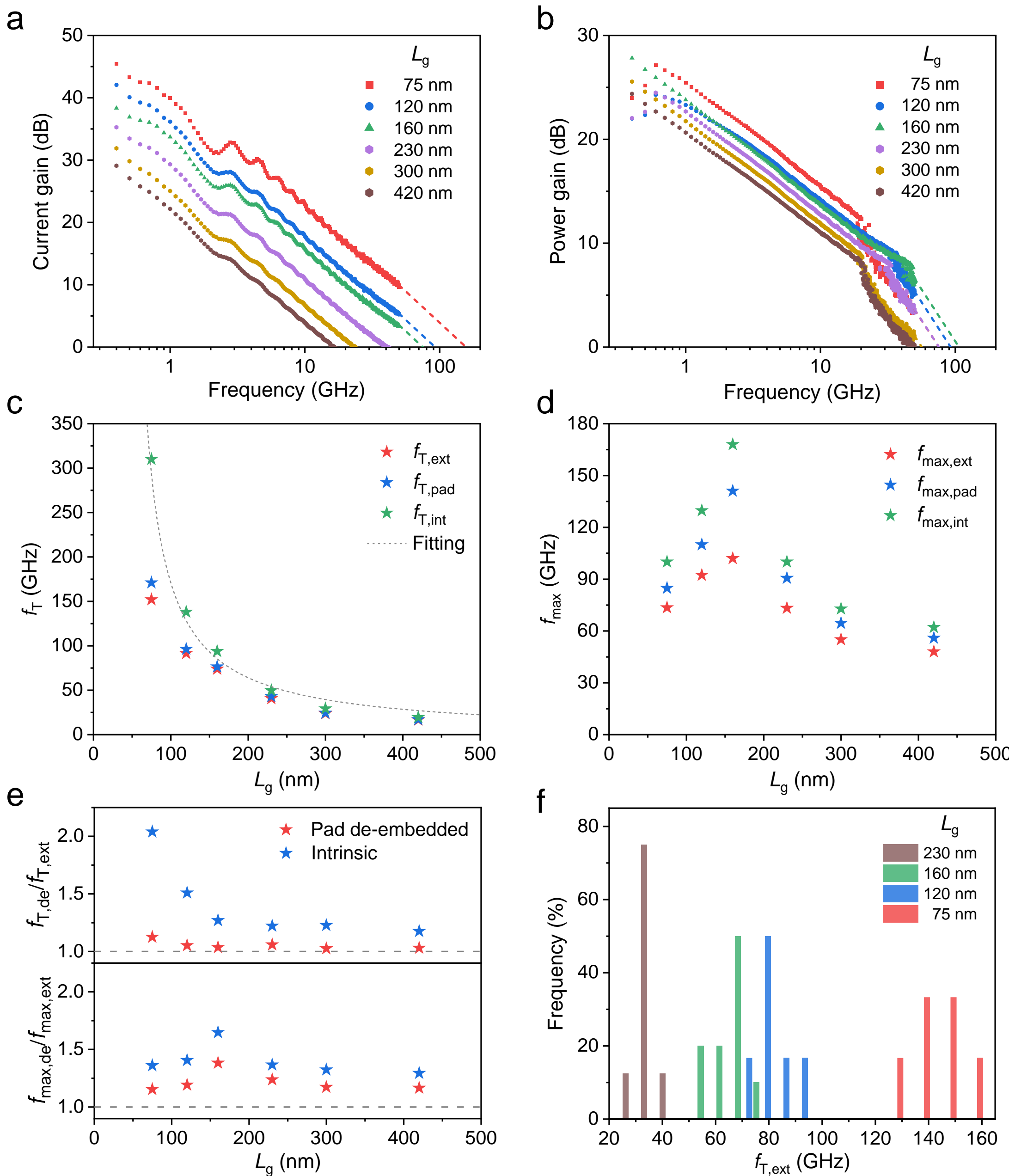

**Fig. 3 | Performance of flexible RF transistors with different $L_g$. a**,**b**, Extrinsic current gain (**a**) and extrinsic power gain (**b**) versus frequency for representative flexible RF transistors with different $L_g$. The slope of the extension line is -20 dB dec$^{-1}$. **c**, Changes in the extrinsic, pad de-embedded, and intrinsic $f_T$ with $L_g$. The $f_{T,int}$ is fitted with an inversely proportional function with $L_g$ ($f_{T,int} \propto 1/L_g$, dashed curve). **d**, Changes in the extrinsic, pad de-embedded, and intrinsic $f_{max}$ with $L_g$. **e**, Pad and intrinsic de-embedding ratios for flexible RF transistors with different $L_g$, with dashed lines representing the de-bedding ratio of 1. **f**, Statistical distributions of $f_{T,ext}$ for flexible RF transistors with different $L_g$.

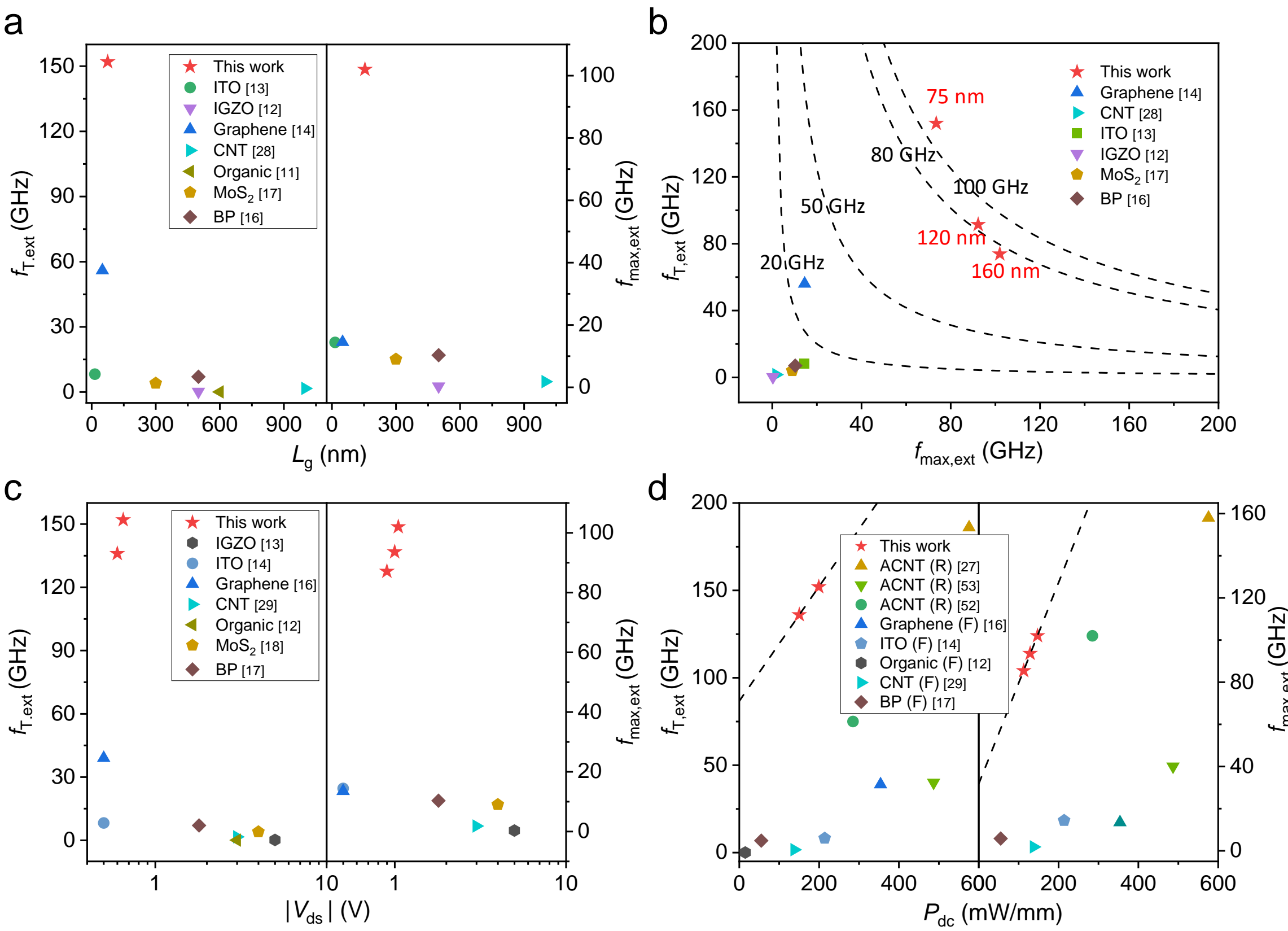

**Fig. 4 | Benchmarking of performance and power consumption.** **a**, Comparison of $f_{\mathrm{T,ext}}$ (left) and $f_{\mathrm{max,ext}}$ (right) versus $L_g$ between our work and other flexible RF MOSFETs based on various materials. **b**, Comparison of $f_{\mathrm{T,ext}}$ versus $f_{\mathrm{max,ext}}$ between our work and other flexible RF MOSFETs. The three data points from our work correspond to flexible RF transistors with $L_g$ = 75 nm, 120 nm, and 160 nm. Dashed curves represent contour lines when the geometric mean, $(f_{\mathrm{T,ext}} \times f_{\mathrm{max,ext}})^{1/2}$, is 20 GHz, 50 GHz, 80 GHz, and 100 GHz. **c**, Comparison of $f_{\mathrm{T,ext}}$ (left) and $f_{\mathrm{max,ext}}$ (right) versus the supply voltage between our work and other flexible RF MOSFETs. **d**, Comparison of $f_{\mathrm{T,ext}}$ (left) and $f_{\mathrm{max,ext}}$ (right) versus $P_{\mathrm{dc}}$ among flexible RF MOSFETs based on different materials and CNT-based RF transistors on rigid substrates. (R) and (F) denote 'rigid' and 'flexible', respectively.

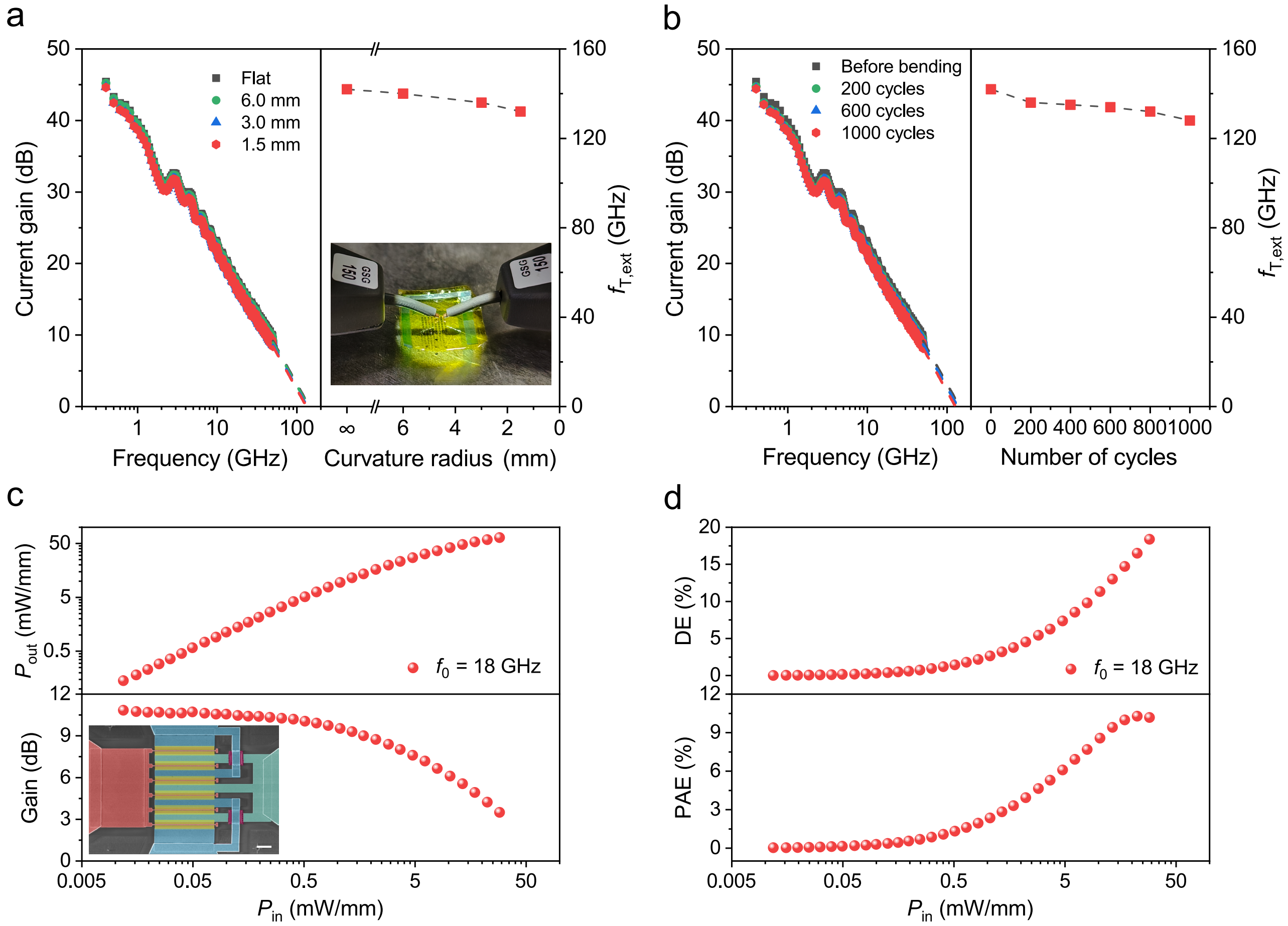

**Fig. 5 | Flexibility test of flexible RF transistors and performance of flexible power amplifiers. a**, Current gain of a typical flexible RF transistor with $L_g$ = 75 nm under different bending radii of curvature (left) and the relationship between $f_{T,ext}$ and the curvature radius (right). Inset: Photograph showing electric characterization of the flexible RF transistors after being released from bending. **b**, Current gain of a typical flexible RF transistor after multiple bending cycles with a curvature radius of 3 mm (left) and the relationship between $f_{T,ext}$ and number of bending cycles (right). **c**, The output power ($P_{out}$, top) and power gain (bottom) versus input power ($P_{in}$) for a flexible RF power amplifier with $L_g$ = 160 nm operating at a $f_0$ of 18 GHz. Bias conditions are $V_{gs}$ = -0.8 V and $V_{ds}$ = -1.0 V. The source and load impedances are 50 Ω and 47.32 + 46.67j Ω, respectively. Inset: SEM image of the flexible RF power amplifier. Scale bar: 3 μm. **d**, The drain efficiency (DE, top) and the power-added efficiency (PAE, bottom) versus $P_{in}$ for a flexible RF power amplifier operating at a $f_0$ of 18 GHz.

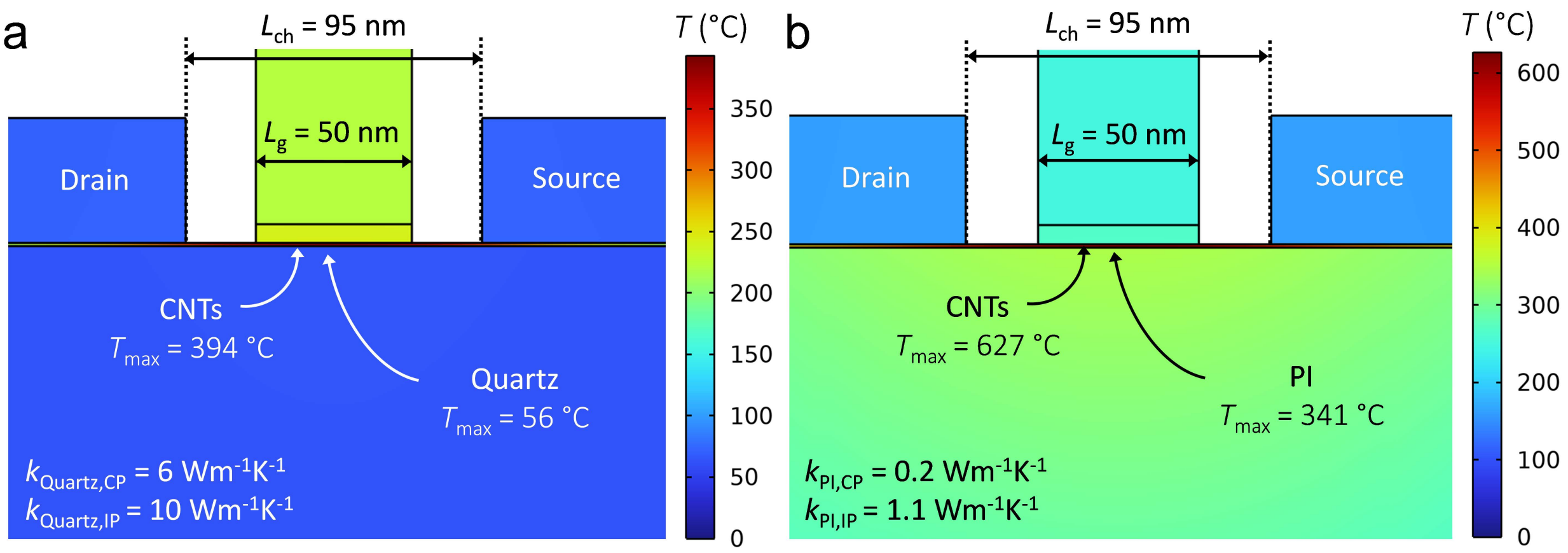

**Extended Data Fig. 1 | Finite-element electro-thermal simulation of an electrically optimized CNT-based RF transistor. a**,**b**, Heat map showing the temperature distribution of a state-of-the-art RF transistor on the rigid substrate (crystal quartz) (**a**) and the flexible substrate (PI) (**b**), with the same power input of 3.36 mW[27].

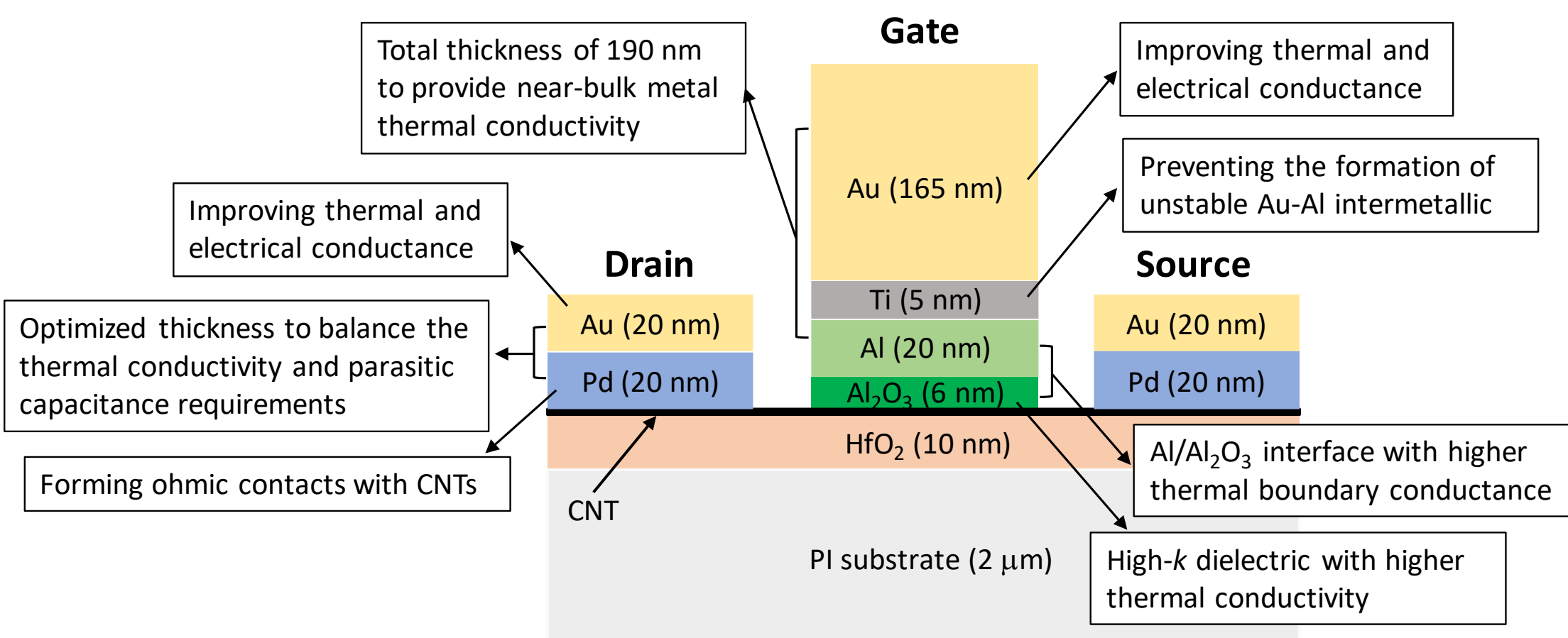

**Extended Data Fig. 2 | Electro-thermal co-design strategy for our RF transistor.** Material selection, interface engineering, and structure optimization are summarized.

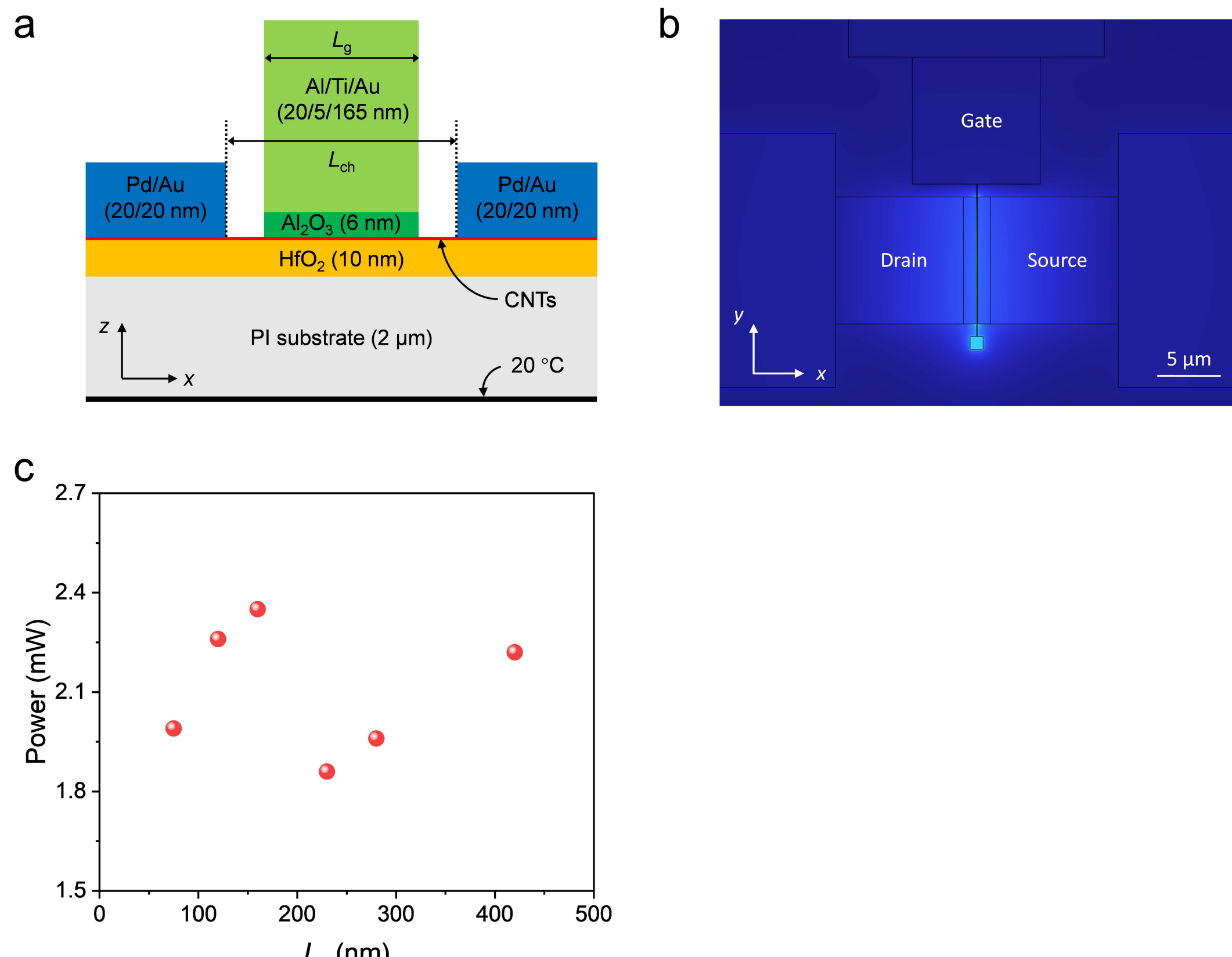

**Extended Data Fig. 3 | Finite-element electro-thermal simulation of flexible CNT-based RF transistors.** **a**, Cross-section schematic of the CNT transistor, showing the dimensions of each segment (not drawn to scale). The bottom of the PI substrate is set to room temperature, 20 °C. **b**, Simulated top-view temperature map of the CNT transistor with $L_{ch}$ = 120 nm, showing negligible temperature rise at the electrode pads. A zoomed-in temperature profile is provided in Fig. 2a,b of the main text. **c**, The power of the flexible RF transistor for different $L_g$.

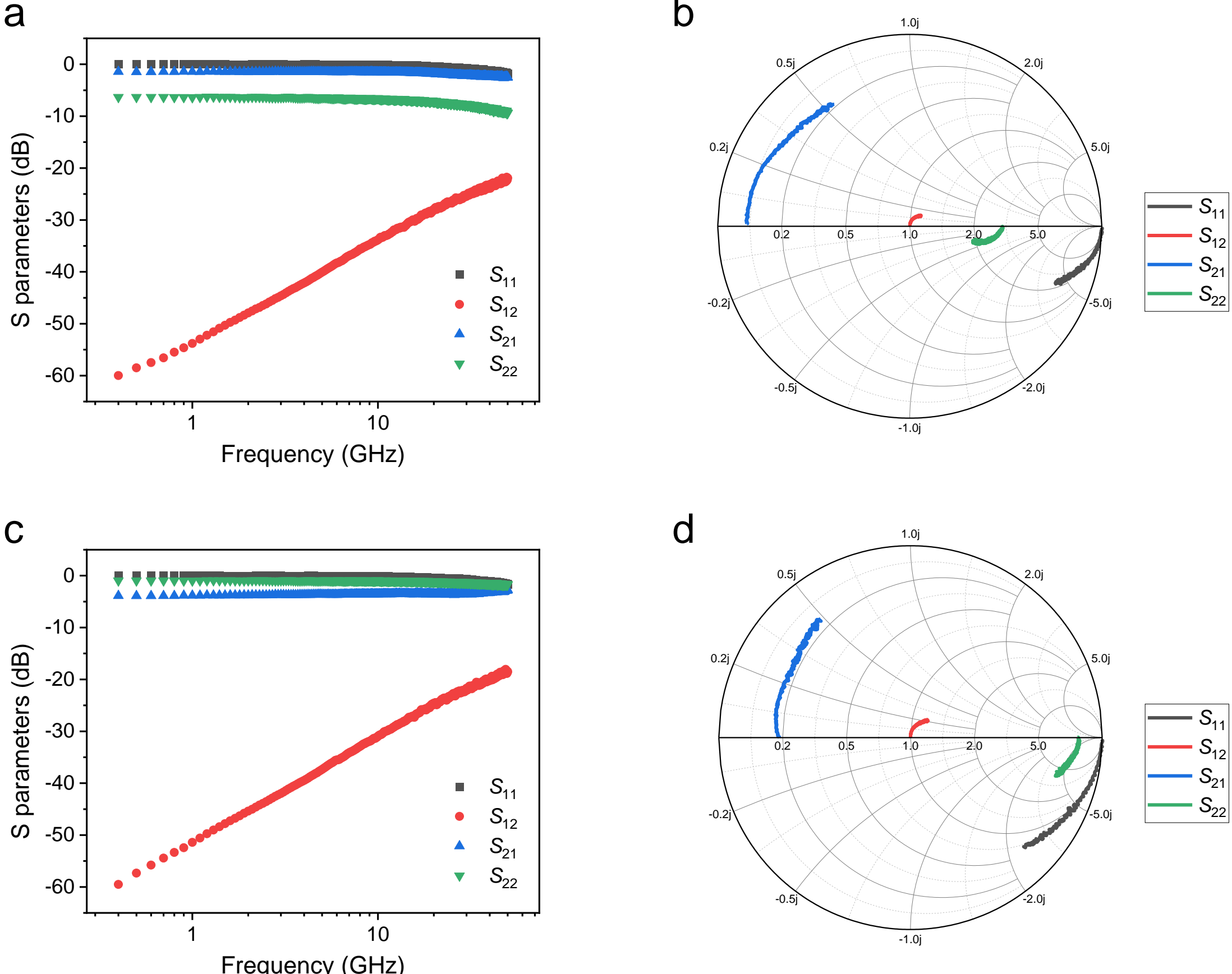

**Extended Data Fig. 4 | S-parameters of flexible CNT-based RF transistors.** **a**, Magnitude of measured extrinsic S-parameters for the flexible RF transistor with $L_g$ = 75 nm as a function of input signal frequency. **b**, Smith chart representation of the S-parameters measured in (**a**). **c**, Magnitude of measured extrinsic S-parameters for the flexible RF transistor with $L_g$ = 160 nm as a function of input signal frequency. **d**, Smith chart representation of the S-parameters measured in (**c**).

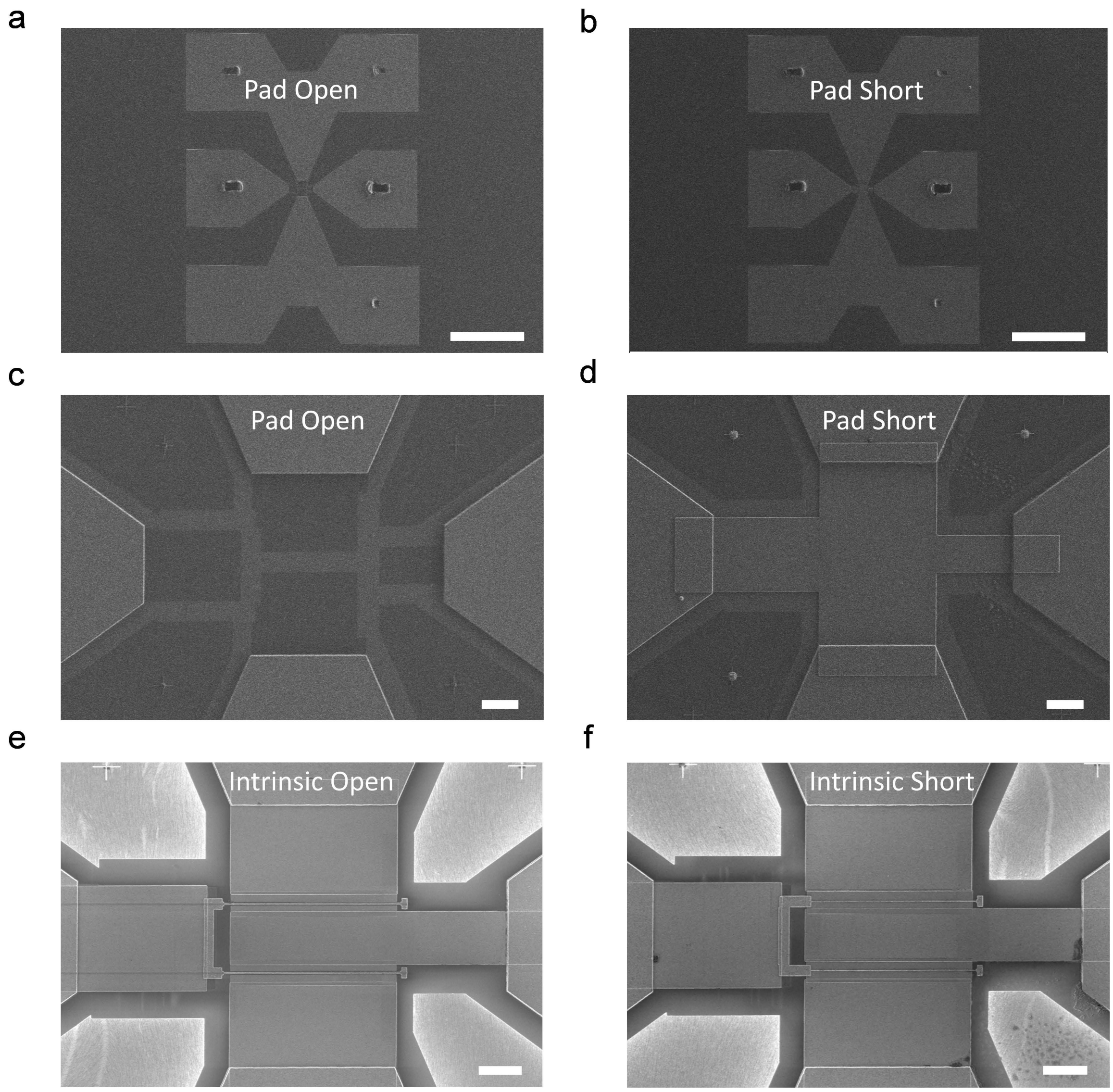

**Extended Data Fig. 5 | SEM images of de-embedding structures.** **a**,**b**, SEM images of the pad de-embedding open (**a**) and short structures (**b**). Scale bar: 100 μm. **c**,**d**, Zoomed-in SEM images of the pad de-embedding open (**c**) and short structures (**d**). Scale bar: 4 μm. **e**, **f**, Intrinsic de-embedding open (**e**) and short structures (**f**). Scale bar: 3 μm.

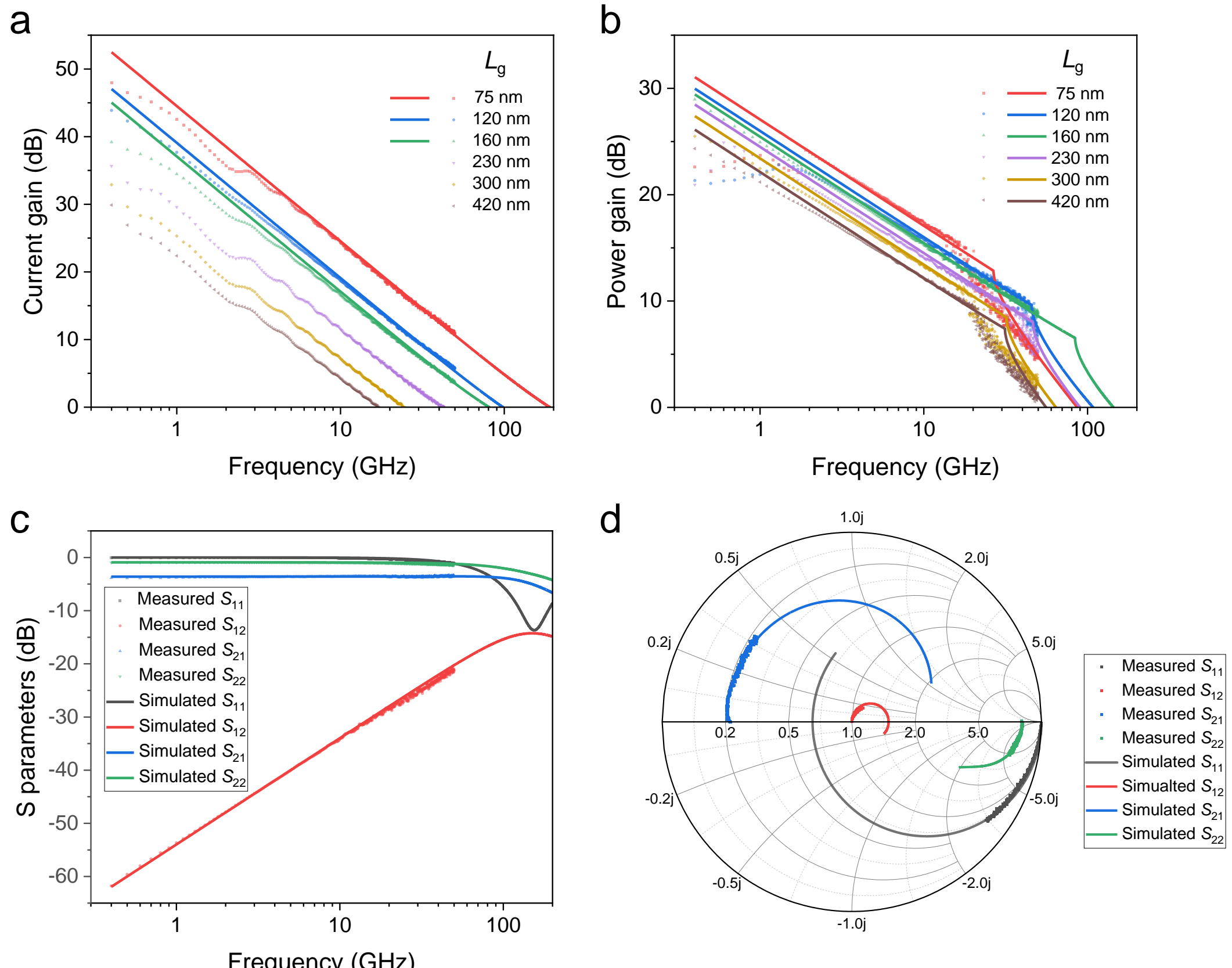

**Extended Data Fig. 6 | Simulation of CNT RF transistors based on AC small-signal equivalent circuit model. a**,**b,** Pad de-embedded current gains (**a**) and power gains (**b**) of flexible RF transistors based on aligned CNTs with different $L_g$ values as a function of input signal frequency from the measured (dot) and simulated (line) S-parameters. **c**, Magnitude of measured (dot) and simulated (line) pad de-embedded S-parameters for the flexible RF transistor with $L_g$ = 160 nm as a function of input signal frequency. The $V_{gs}$ and $V_{ds}$ are -0.4 V and -1.05 V, respectively. **d**, Smith chart representation of the S-parameters measured and simulated in (**c**). The small signal equivalent circuit model simulation was conducted with Advanced Design System (ADS 2020, Keysight technologies), as shown in Supplementary Fig. 5. The simulation frequency is from 400 MHz to 200 GHz with a step of 100 MHz.

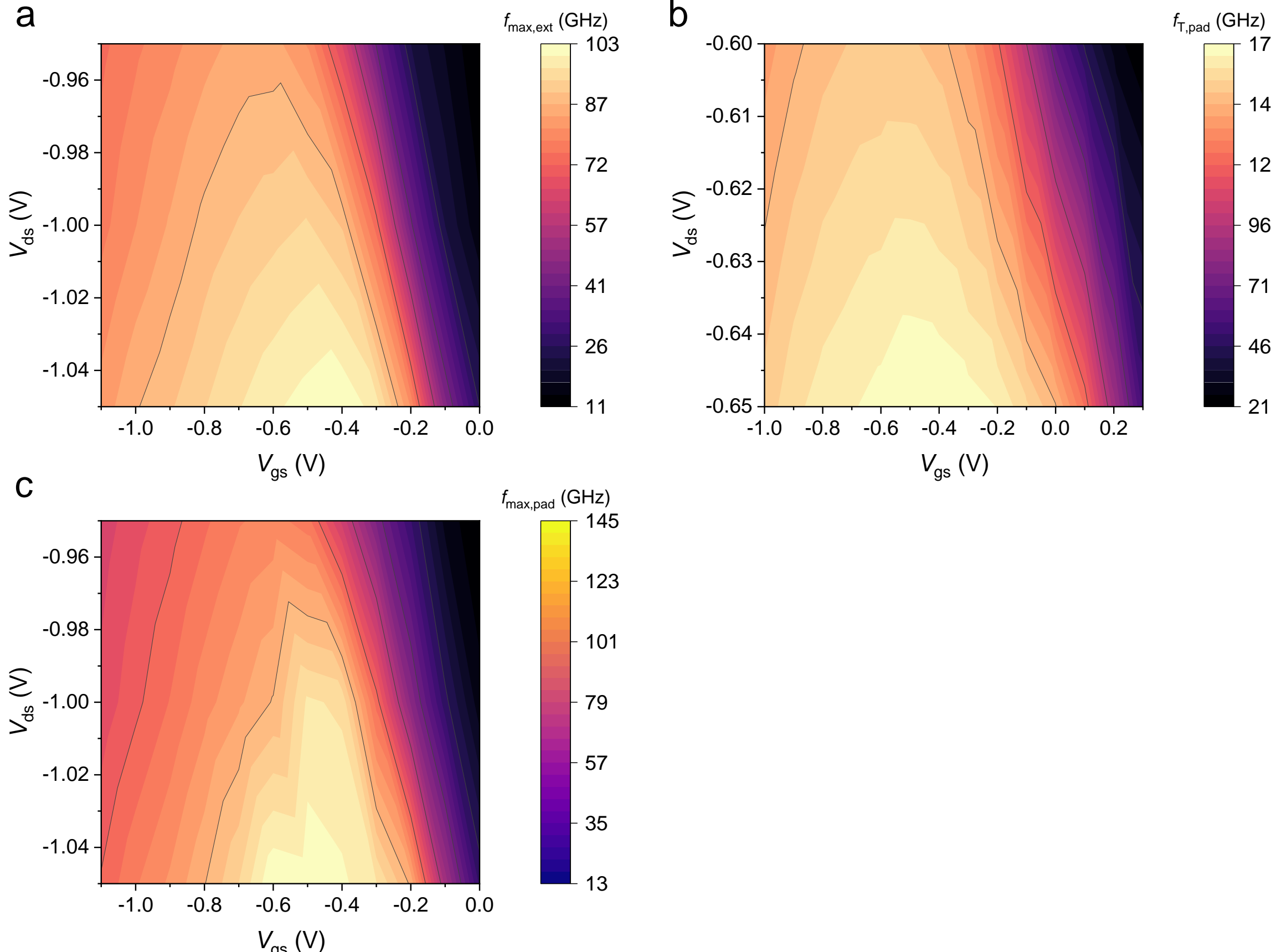

**Extended Data Fig. 7 | Bias conditions to achieve the highest $f_{max,ext}$, $f_{T,pad}$, $f_{max,pad}$. a**, Bias conditions ($V_{gs}$ and $V_{ds}$) of $f_{max,ext}$ for the flexible RF transistor with $L_g$ = 160 nm. **b**, Bias conditions ($V_{gs}$ and $V_{ds}$) of $f_{T,pad}$ for the flexible RF transistor with $L_g$ = 75 nm. **c**, Bias conditions ($V_{gs}$ and $V_{ds}$) of $f_{max,pad}$ for the flexible RF transistor with $L_g$ = 160 nm.

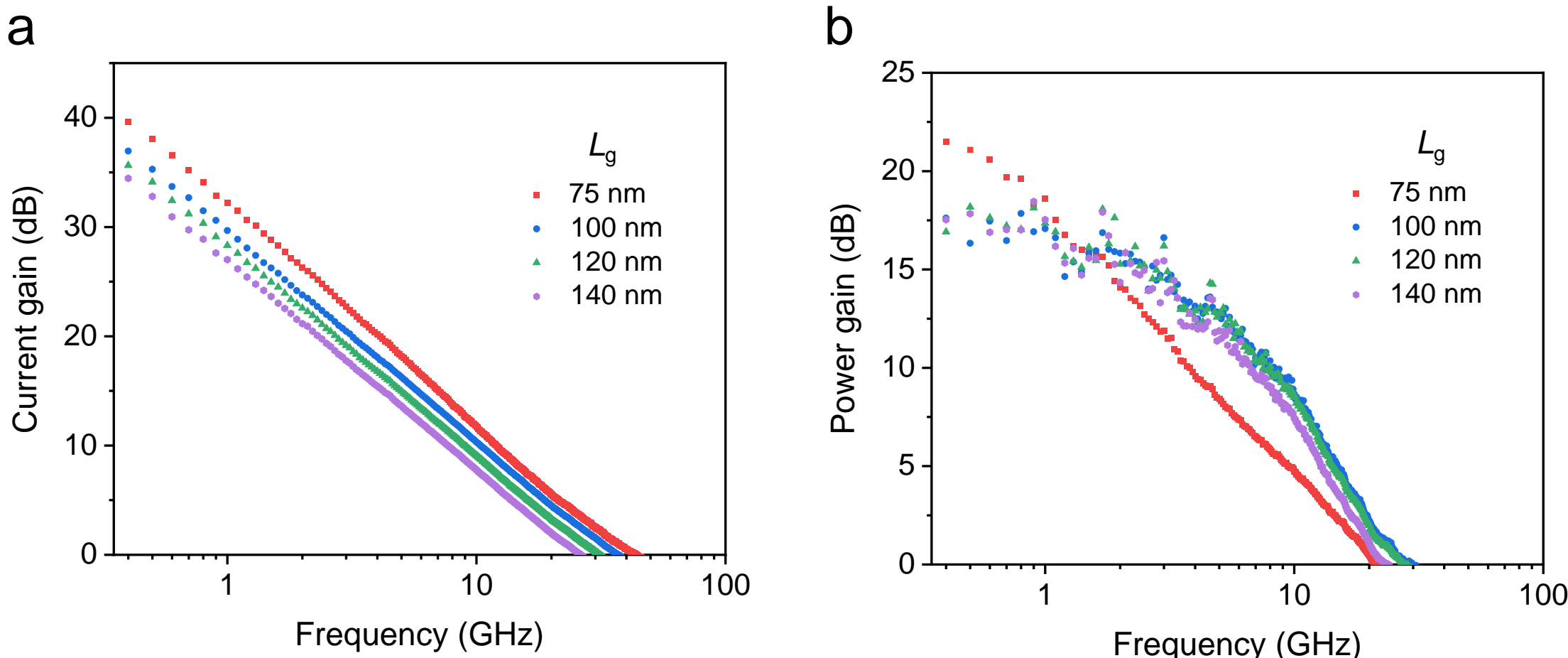

**Extended Data Fig. 8 | RF performance of flexible CNT-based RF transistors fabricated without electro-thermal co-design. a**,**b**, Extrinsic current gain (a) and power gain (b) vs. frequency for transistors with different $L_g$. These devices exhibit significantly reduced performance, with maximum $f_{T,ext}$ of 44 GHz ($L_g$ = 75 nm) and $f_{max,ext}$ of 30 GHz ($L_g$ = 100 nm). Additionally, more than half of these devices suffered irreversible breakdown immediately following RF measurements (see SEM images in Supplementary Fig. 7, highlighting their limited reliability and single-use nature. In contrast, flexible RF transistors with electro-thermal co-design demonstrated consistently repeatable high performance, emphasizing the critical role of effective thermal management.

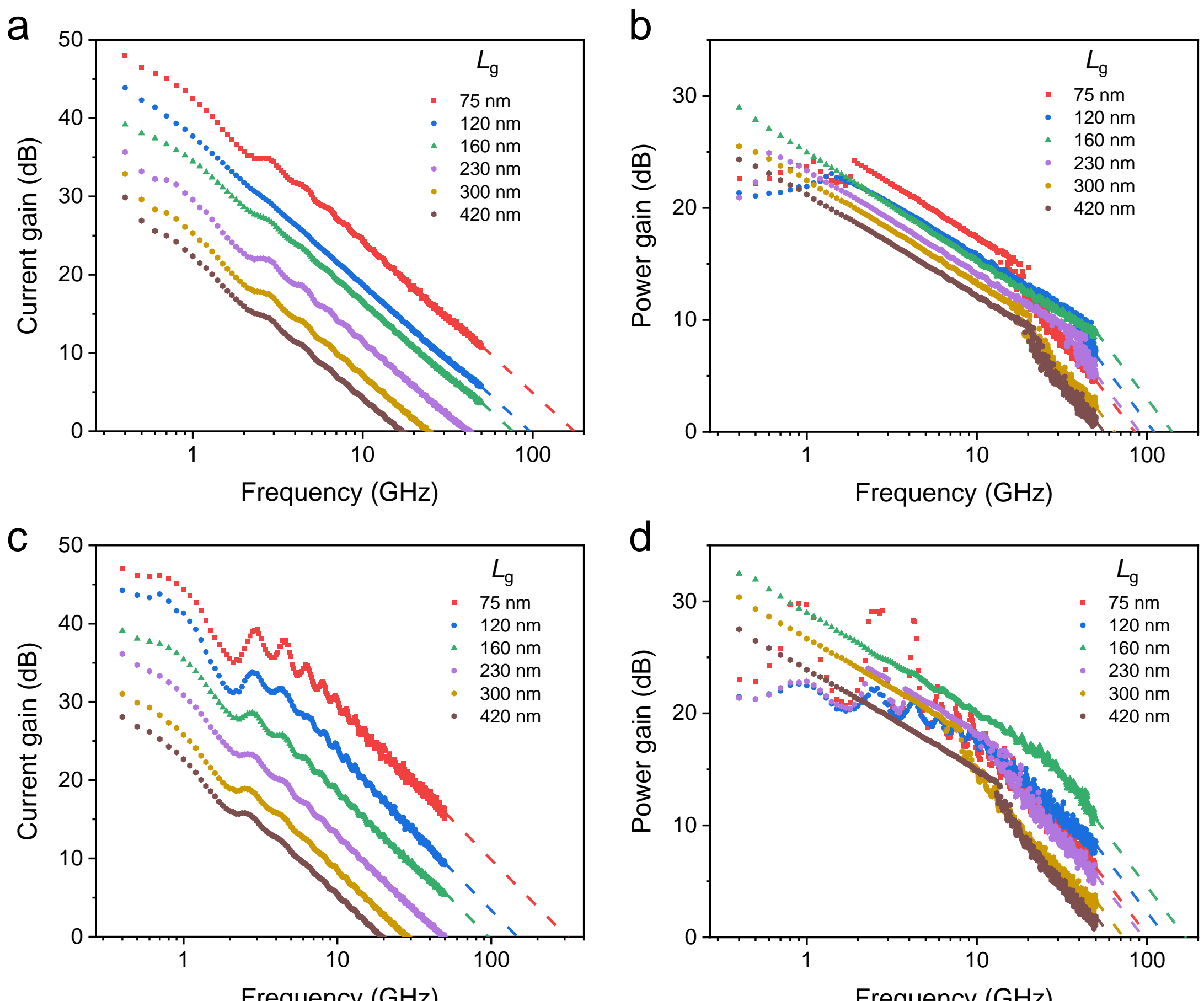

**Extended Data Fig. 9 | Pad de-embedded and intrinsic performance of flexible RF transistors with different $L_g$. a,b,** Pad de-embedded current gains (**a**) and power gains (**b**) of flexible RF transistors based on aligned CNTs with different $L_g$ values as a function of input signal frequency from the S-parameter measurements. **c,d**, Intrinsic de-embedded current gains (**c**) and power gains (**d**) of flexible RF transistors with different $L_g$ values. The input power for the small signal is -10 dBm. The upper-frequency limit of the measurement instrument is 50 GHz, and the lines beyond 50 GHz were extrapolated with a slope of -20 dB per decade.

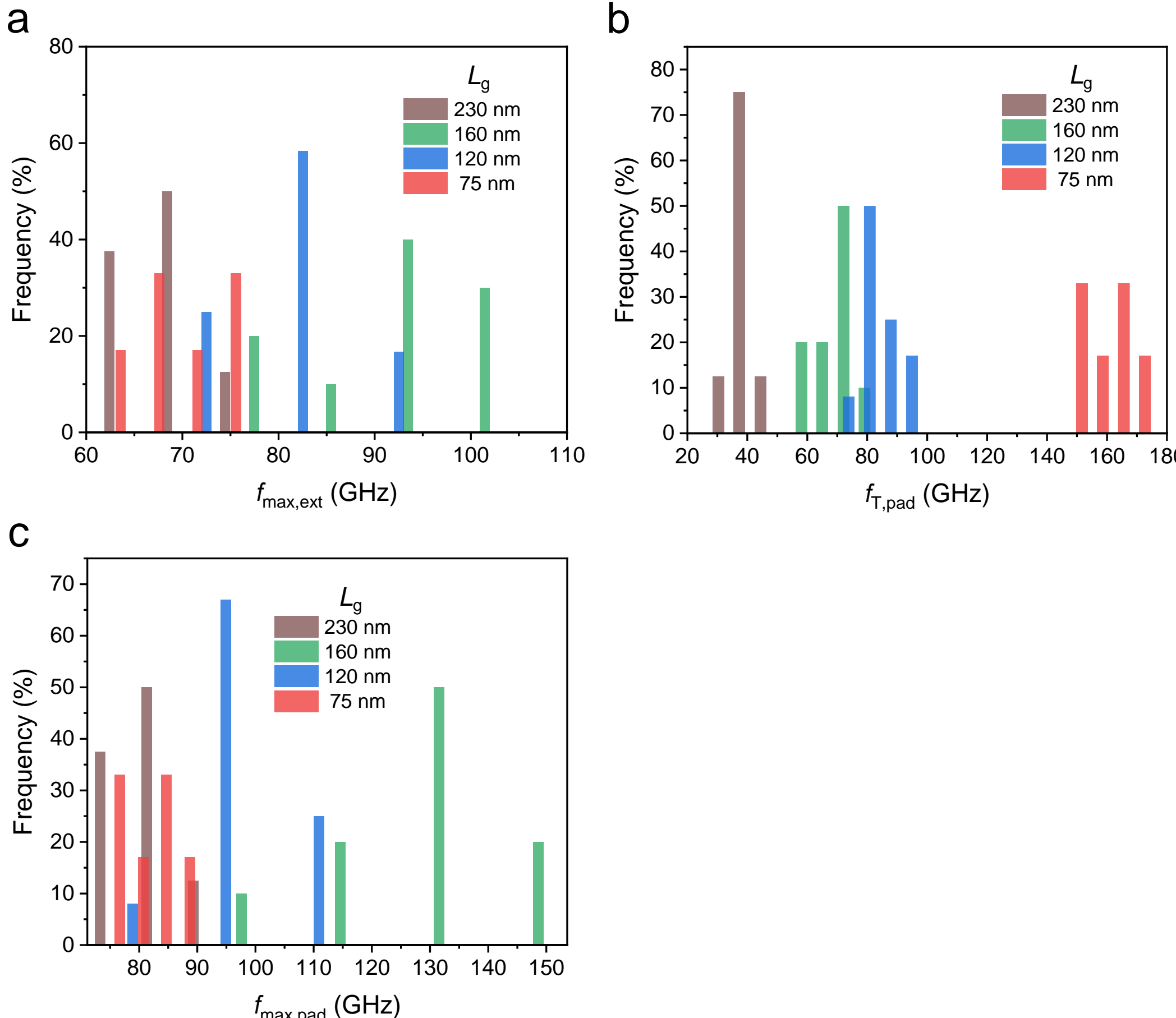

**Extended Data Fig. 10 | Statistical distributions of the $f_{max,ext}$, $f_{T,pad}$ and $f_{max,pad}$. a**,**b**,**c**, Statistical distributions of $f_{max,ext}$ (**a**), $f_{T,pad}$ (**b**) and $f_{max, pad}$ (**c**) for 12 flexible RF transistors, each with the $L_g$ of 75 nm, 120 nm, 160 nm and 230 nm. The average values and standard deviations are summarized in Supplementary Table 6, highlighting the excellent uniformity of the devices.

**Supplementary Note 1 | Influence of key design parameters on thermal behavior and RF performance**

To provide more quantitative guidance for device design, we conducted electro-thermal simulations using Advanced Design System (ADS) and COMSOL Multiphysics. The results, summarized in Supplementary Table 2, explicitly elucidate the trade-offs between thermal management and RF performance in flexible CNT RF transistors.

**Source/drain electrode thickness.** We first examine the influence of source/drain electrode thicknesses. Increasing the metal thickness enhances heat dissipation by improving thermal spreading, but simultaneously degrades RF performance due to increased parasitic capacitance. Within this trade-off space, a 20/20 nm Pd/Au stack provides an optimal balance between heat dissipation and RF performance. Thinner electrodes (10/10 nm) lead to excessive self-heating, while thicker electrodes (40/40 nm) cause pronounced RF degradation with diminishing thermal benefit.

**Gate electrode thickness.** We also evaluate the impact of gate electrode thicknesses. Increasing the gate thickness improves both $f_{max}$ and heat dissipation, owing to reduced gate resistance ($R_g$) and enhanced thermal conductivity of the gate metal. In practice, however, fabrication constraints on flexible substrates limit the achievable gate thickness. We therefore chose a gate thickness of 190 nm, which approaches the bulk thermal conductivity of gold[45] and provides near-maximal cooling capability while maintaining high RF performance.

**$L_g/L_{ch}$ ratio.** Finally, we investigate the effect of the $L_g/L_{ch}$ ratio, which represents the fraction of the channel length covered by the gate electrode. We found that variations in this ratio have a minimal impact on heat dissipation, resulting in similar peak device temperatures, but significantly influence RF performance. An optimal ratio of $L_g/L_{ch}$ = 0.625 ($L_g$ = 75 nm, $L_{ch}$ = 120 nm) yields the highest RF performance while maintaining a relatively low device temperature.

**Supplementary Note 2 | Finite-element electro-thermal modeling**

We estimate the temperature rise ($\Delta T$, above room temperature) in our CNT transistors using finite-element electro-thermal modeling[37]. Devices with varying channel lengths and operating characteristics, as listed in Supplementary Table 7 below, were modeled to estimate their peak $\Delta T$. These specific devices and operating conditions were simulated at the power levels that devices experience during our RF measurements, as discussed in the main text. We conducted the steady-state simulations of CNT transistors with dimensions shown in Extended Data Fig. 3a. The bottom of the 2 μm thick PI substrate is taken as the thermal ground (fixed at 20 °C), because it was attached to a metal chuck during our RF measurement. (We tested the simulation with an interfacial thermal conductance between the PI and metal chuck[37,55-57], $h_c \approx 10^5$ $Wm^{-2}K^{-1}$, but found this has little impact on the simulated peak $\Delta T$.)

The thermal conductivity (TC) of the materials and thermal boundary conductance (TBC) at material interfaces used in our simulations are given in Supplementary Table 8. Significant variations in CNT TC have been reported in the literature[58-60]. For example, the TC of CNTs depends on factors such as their quality, diameter, length, density, and the substrate on which they are sitting. For our dense (~150 CNTs/μm) aligned CNTs on $HfO_2$, the TC of such arrays remains largely unknown. Moreover, predicting the TC of CNTs using the Wiedemann-Franz law[61,62] is challenging, as phonons, rather than electrons, dominate heat conduction along CNTs and their contacts across all temperatures of practical interest[44]. Consequently, we used the TC values from the theoretical work by Volkov *et al.*[63] as a guideline when estimating the TC of our CNT arrays.

To estimate the peak $\Delta T$ in our devices, we also need to determine the electrical power distribution across the device. Despite the shortest channel length being 120 nm, our devices are not in the quasi-ballistic regime because the average total resistance per CNT is ≈ 81.3 kΩ (~54.2 Ω device total resistance extracted from Fig. 1e and for ~1,500 CNTs in parallel), significantly higher than the quantum ballistic transport limit of single CNT, with ~6.5 kΩ resistance[64]. Thus, we model the CNT arrays with uniform sheet resistance across their length. Pd metal is used to form high-quality ohmic contacts at the source and drain, to minimize contact resistance[33,64-66]. In Extended Data Fig. 3b, the simulated top view of the $\Delta T$ distribution for a CNT transistor with $L_{ch}$ = 120 nm shows minimal temperature rise of the big metal Source/Drain and Gate pads.

To understand the temperature profile in our devices, we estimate the thermal healing length (i.e., the characteristic temperature decay length) along the CNT channel, written generally as[67]:

$$L_{\mathrm{H}} = \sqrt{\frac{k_{\mathrm{film}} d_{\mathrm{film}} d_{\mathrm{substrate}}}{k_{\mathrm{substrate}}}},$$

where $k_{\mathrm{film}}$ and $d_{\mathrm{film}}$ are the thermal conductivity and thickness of the heater element (here, the CNTs), while $k_{\mathrm{substrate}}$ and $d_{\mathrm{substrate}}$ are the thermal conductivity and thickness of the substrate under the heater. To make this more specific to the CNT devices, we can rewrite their $L_{\mathrm{H}}$ as:

$$L_{H} = \sqrt{k_{\mathrm{CNT,In\text{-}plane}} d_{\mathrm{CNT}} R_{\mathrm{th}}},$$

where $R_{\mathrm{th}}$ is the thermal resistance between CNT and heat sink at the bottom of the PI:

$$R_{\mathrm{th}} = \frac{1}{G_{\mathrm{CNT\text{-}PI}}} + R_{\mathrm{PI}}.$$

Here $R_{\mathrm{PI}}$ is the thermal resistance of the PI substrate, accounting for the shape factor of a circular disk heater on a semi-infinite substrate[49,68], which has also been found as a good approximation for a rectangular heater (i.e., the shape of the transistor channel):

$$R_{\mathrm{PI}} \approx \frac{WL}{2k_{\mathrm{PI,eff}}},$$

where $k_{\mathrm{PI,eff}} \approx (k_{\mathrm{PI,cross\text{-}plane}} \times k_{\mathrm{PI,In\text{-}plane}})^{1/2}$, and $G_{\mathrm{CNT\text{-}PI}}$ is the total thermal conductance between the channel of CNTs and PI substrate, including the TBC at CNT-$HfO_2$ and $HfO_2$-PI interfaces, and the thermal conductance of the $HfO_2$ insulator [see Extended Data Fig. 3a]:

$$G_{\mathrm{CNT\text{-}PI}} = \left( \frac{1}{TBC_{\mathrm{CNT\text{-}HfO_2}}} + \frac{1}{TBC_{\mathrm{HfO_2\text{-}PI}}} + \frac{d_{\mathrm{HfO_2}}}{k_{\mathrm{HfO_2}}} \right)^{-1}.$$

With TC and TBC values in Supplementary Table 8, using $L_{ch}$ of 120 nm, we estimate $L_{\mathrm{H}}$ ≈ 283 nm along the CNT channel. This closely matches our simulation results, which show $L_{\mathrm{H}}$ ≈ 300 nm, as shown in Fig. 2b of the main text.

We also conducted the sensitivity analysis of device temperature with respect to various material properties. Increasing (i.e., doubling) the in-plane TC of CNTs to 100 $Wm^{-1}K^{-1}$ resulted in a ~20% drop in the maximum CNT channel temperature. Additionally, to address uncertainties in TBCs at various material interfaces, we adopted conservative assumptions using lower-bound estimates (as detailed in Supplementary Table 8) to model the worst-case scenario. Our approach thus provides an upper-bound estimate for the peak device temperature.

**Supplementary Note 3 | Limitations of direct thermal measurements and experimental validation of electro-thermal simulations**

We attempted Raman thermometry and evaluated several established thermal measurement techniques to experimentally probe the CNT channel temperature. However, owing to the specific device geometry and nanoscale channel dimensions, direct measurement of the true CNT channel temperature presents inherent experimental challenges.

As shown in Supplementary Fig. 4a,b, no discernible CNT Raman peaks were detected from our RF devices ($L_g$ = 75 nm and 420 nm), which we attribute to the top-gated architecture in which the thick metal/dielectric gate stack strongly attenuates the Raman signal from the underlying CNTs. In contrast, clear Raman signatures were observed from blank CNT films on both Si and flexible PI substrates, confirming that the signal loss originates from the device geometry rather than the CNT material.

To circumvent this limitation, we performed Raman thermometry[68-70] on bottom-gated CNT transistors fabricated on flexible substrates with a channel length of 200 nm. Raman spectra were acquired with and without electrical bias (Supplementary Fig. 4c). Although current-induced Raman peak shifts are observable, they reflect a spatially averaged temperature over all CNTs within the micrometer-scale laser spot, including both electrically heated CNTs in the channel and much cooler CNTs outside the contact region. Because the probing area is much larger than the channel length, localized Joule heating is averaged out, and the extracted Raman temperature can substantially underestimate the true hotspot temperature in the short-channel region.

Nevertheless, the Raman signal still enables an estimate of the temperature rise, albeit a lower bound of the true channel heating. A calibration was established by measuring CNT films on a heating stage at controlled temperatures (Supplementary Fig. 4d), yielding the relationship between temperature and Raman peak position. Using this calibration, the Raman shifts measured under electrical bias (Supplementary Fig. 4c) correspond to a temperature increase of approximately $\Delta T \approx 160$ °C. Even with this underestimation, the extracted $\Delta T$ remains substantial, indicating significant Joule heating in the CNT channel.

We also evaluated several alternative thermal measurement approaches but found that none can directly resolve the CNT channel temperature in our top-gated geometry. Infrared (IR) thermography[71,72] and other thermoreflectance imaging techniques[73,74] are limited by micrometer-scale spatial resolution and, in top-gated devices, primarily capture the gate surface temperature rather than the CNT channel temperature. Although scanning thermal microscopy (SThM)[75,76] and scanning Joule expansion microscopy (SJEM)[77,78] can achieve sub-100-nm spatial resolution, they also probe only the top surface. In our configuration, the top-gate electrode screens the underlying CNTs, preventing direct access to the channel temperature.

To further validate our electro-thermal simulations, we performed electrical measurements on the same bottom-gated CNT device used for Raman thermometry, with 200 nm channel length. CNT breakdown in air is well established to be dominated by oxidation at elevated temperatures, with a reported[41,44,79,80] breakdown temperature of 600 ± 100 °C. In bottom-gated devices, the CNT channel is fully exposed to air, allowing this estimate to be directly applied, whereas in top-gated devices the breakdown temperature is less defined due to the buried channel region[80-82].

Experimentally, the breakdown occurs at a current of nearly 10 mA (1 mA/μm) under $V_{ds}$ = -0.7 V. Finite-element electro-thermal simulations using the device geometry shown in Supplementary Fig. 4e and the thermal parameters listed in Supplementary Table 8 yield a peak temperature rise $\Delta T \approx 629$ °C (Supplementary Fig. 4f), consistent with the reported CNT breakdown temperatures. Taken together, these breakdown measurements and simulations provide an experimentally-anchored validation of our electro-thermal modeling and support the simulated temperature distributions presented for the top-gated devices in Fig. 2a–c of the main text.

**Supplementary Note 4 | De-embedding procedure of flexible RF transistors**

A standard open-short de-embedding method was employed for pad and intrinsic de-embedding. The process is outlined as follows:

(1) Measure the S-parameters of the device under test (DUT), open structure, and short structure.

(2) Convert the measured S-parameters ($S_{DUT}$, $S_{Open}$, $S_{Short}$) to Y-parameters ($Y_{DUT}$, $Y_{Open}$, $Y_{Short}$).

(3) Calculate the de-embedded Y-parameters $Y_{de\text{-}embed}$ using the following equation:

$$\frac{1}{Y_{\text{de-embed}}} = \frac{1}{Y_{\text{DUT}} - Y_{\text{Open}}} - \frac{1}{Y_{\text{Short}} - Y_{\text{Open}}}.$$

(4) Convert the de-embedded Y-parameters $Y_{de\text{-}embed}$ back to de-embedded S-parameters $S_{de\text{-}embed}$.

(5) Calculate the de-embedded $f_T$ and $f_{max}$ using the de-embedded S-parameters $S_{de\text{-}embed}$.

Two types of de-embedding open and short structures were used: pad de-embedding structures and intrinsic de-embedding structures. The pad de-embedding structures, shown in Extended Data Fig. 5a,b (with zoomed-in views in Extended Data Fig. 5c,d, respectively), consist solely of the pads of the flexible RF transistor. In contrast, the intrinsic de-embedding structures, illustrated in Extended Data Fig. 5e,f, include the connection wires and electrodes of the flexible RF transistor. The S-parameters of the pad de-embedding short and open structure are shown in Supplementary Fig. 11 as Smith charts. The $S_{11}$ and $S_{22}$ of the open structure are located near the infinite point of impedance, indicating minimal parasitic effects from the substrate and the measurement system.

In practical applications, the external parasitic elements differ from the laboratory setup. As a result, the circuit-level $f_{T,ext}$ and $f_{max,ext}$ may vary depending on the quality of interconnects, wire bonding, encapsulation, vertical interconnect access, and overall system implementation. By contrast, the pad-de-embedded $f_{T,pad}$ and $f_{max,pad}$ remove pad- and interconnect-related parasitics and are expected to remain essentially unchanged across different system environments, thereby serving as device-level figures of merit that reflect the intrinsic capability of the transistor.

**Supplementary Note 5 | Extraction of $R_g$ in flexible RF transistors with different $L_g$ and its impact on $f_{max}$**

To elucidate the physical origin of the observed $f_{max}$ peak, the gate resistance ($R_g$) of flexible RF transistors in Fig. 3a, b with different channel lengths ($L_g$) was extracted and its influence on RF performance was evaluated. The initial extraction was performed using the small-signal equivalent model (Supplementary Fig. 5) implemented in Advanced Design System (ADS 2020; Keysight Technologies), and the results are summarized in Supplementary Fig. 10a. The analysis indicates that $R_g$ increases markedly as $L_g$ is scaled below 160 nm, and the extracted $R_g$ reaches 116 Ω with $L_g$ = 75 nm.

To validate these results, $R_g$ was additionally obtained from the Z-parameters formulations based on the small-signal equivalent circuit model[83]. The extracted values show good agreement between the two methods (Supplementary Fig. 10b and Supplementary Table 9), confirming the reliability of the extraction.

Based on the small-signal model, $f_{max}$ is expressed as[51]

$$f_{\text{max}} = \frac{f_{\text{T}}}{2[g_{\text{d}}(R_{\text{s}} + R_{\text{g}}) + 2\pi f_{\text{T}} C_{\text{gd,p}} R_{\text{g}}]^{1/2}},$$

where $g_d$, $R_s$ and $C_{gd,p}$ represents the drain conductance, source series resistance and parasitic gate-to-drain capacitance, respectively. In a first-order approximation, the first term in the denominator becomes negligible, yielding

$$f_{max} \propto (f_T/R_g)^{1/2}.$$

Although $f_T$ increases with channel-length scaling (Fig. 3c), the rapid rise in $R_g$ at short $L_g$ introduces a competing effect, leading to the emergence of a maximum in $f_{max}$. This interpretation is consistent with the experimental trend, as plotting $(f_T/R_g)^{1/2}$ yields a peak coincident with the measured $f_{max}$ behavior (Supplementary Fig. 10c).

To further illustrate the critical role of $R_g$, a hypothetical case was considered in which $R_g$ varies more gradually. Using devices with $L_g \geq 160$ nm, a linear fit was performed to extrapolate $R_g$ for shorter channels, followed by recalculation of $(f_T/R_g)^{1/2}$ . Under this condition, $(f_T/R_g)^{1/2}$ increases monotonically with scaling (Supplementary Fig. 10d), eliminating the peak.

Together, these results demonstrate that the experimentally observed $f_{max}$ peak originates from the rapid increase in $R_g$ at short channel lengths, highlighting the importance of suppressing $R_g$ to further advance the RF performance of flexible CNT transistors.

**Supplementary Note 6 | Roadmap for performance improvements of flexible RF transistors**

In this work, the highest pad-de-embedded cutoff and maximum oscillation frequencies achieved by flexible RF transistors based on aligned CNT arrays are **171/141 GHz** ($f_{T,pad}/f_{max,pad}$) **at $L_g$ = 75 nm and 160 nm**, respectively. By comparison, CNT RF transistors on rigid substrates using the same kind of aligned CNT arrays and comparable channel length have reported **161/144 GHz at $L_g$ = 90 nm and 130 nm**[27]. With further channel-length scaling, device performance on rigid substrates has reached **345/192 GHz at $L_g$ = 50 nm**[27], and even **551/1024 GHz at $L_g$ = 35 nm** with highly optimized gate engineering[84].

We attribute the remaining performance gap between flexible and rigid CNT RF transistors to the following factors:

1. **Thermal and mechanical degradation.** When device dimensions are relatively large, the performance difference between flexible and rigid platforms is limited. However, as $L_g$ is further scaled down, device thermal resistance increases, and self-heating in flexible devices becomes increasingly pronounced due to the limited heat dissipation associated with the low thermal conductivity of polymer substrates, leading to more severe mobility degradation and reduced transconductance in flexible devices. In addition, cyclic bending and accumulated mechanical strain can induce mechano–electrical coupling effects that compromise long-term stability and RF performance.

2. **Limitations in lithographic precision for sub-100 nm $L_g$ devices on flexible substrates.** Unlike rigid wafers, flexible substrates inherently suffer from mechanical deformation, surface non-planarity, and dimensional instability during fabrication; moreover, surface charging on insulating polymers hinders high-resolution e-beam lithography. Together, these factors make sub-100-nm patterning and alignment significantly more challenging, thereby limiting aggressive gate scaling and the realization of ultra-short, precisely defined channels.

3. **Challenges in implementing highly optimized gate structures for high-frequency operation.** State-of-the-art rigid CNT RF devices benefit from sophisticated gate architectures (e.g., T-shaped gates, Y-shaped gates, and buried gates) that effectively reduce $R_g$ and parasitic capacitances. Translating such architectures onto flexible substrates remains technically demanding due to process compatibility, mechanical robustness requirements, and integration constraints.

4. **Variations in aligned CNT array quality and density on flexible substrates.** Compared with rigid substrates, flexible substrates typically present differences in surface chemistry, roughness, and thermal budget, making direct CNT alignment, assembly, and integration more challenging. Transfer and printing processes can also introduce defects, misalignment, and local density fluctuations, which in turn limit achievable $g_m$ and RF performance.

Overall, our strategy for closing the performance gap between flexible and rigid CNT RF transistors follows a stepwise and principle-driven approach. First, **thermal and mechanical robustness must be prioritized**, as excessive self-heating on polymer substrates and mechano–electrical coupling under bending can otherwise offset the benefits of channel scaling and advanced gate engineering. Establishing efficient heat-spreading pathways, operating near the mechanical neutral plane, and defining safe bending and power-operation windows are therefore essential prerequisites.

Second, **aggressive channel-length scaling and gate-structure optimization should be pursued only after the thermo-mechanical environment is brought under control**. In this regime, improving lithographic precision on mechanically compliant substrates and implementing low-$R_g$, high-frequency gate architectures can effectively push $f_T$ and $f_{max}$ toward the levels of rigid devices.

Third, **further gains ultimately depend on the intrinsic capability of the active material**, namely the density and uniformity of aligned CNT arrays on flexible substrates. Continued advances in CNT assembly and transfer processes, together with electro–thermal–mechanical co-design, are therefore expected to lift transconductance and enable further RF performance scaling.

Following these design principles, we propose the quantitative roadmap illustrated in Supplementary Fig. 15 (simulation parameters used in ADS simulations summarized in Supplementary Table 10):

- **Step 0:** Realize high-quality flexible substrates with low surface roughness, high thermal conductivity, and low RF loss, establishing the thermo-mechanical foundation.
- **Step 1:** Introduce optimized gate architectures to reduce $R_g$ to ~10 Ω. improving $(f_{T,pad} \times f_{max,pad})^{1/2}$ to 227 GHz.
- **Step 2:** Scale $L_g$ while maintaining thermal and mechanical stability through co-design. Specifically,

  **Step 2-1**: at $L_g$ = 50 nm, $(f_{T,pad} \times f_{max,pad})^{1/2}$ can reach 297 GHz;

  **Step 2-2**: at $L_g$ = 30 nm, $(f_{T,pad} \times f_{max,pad})^{1/2}$ can reach 377 GHz.
- **Step 3:** Enhance aligned CNT array quality and density on flexible substrates to increase $g_m$ to 1.5 mS/μm, enabling $(f_{T,pad} \times f_{max,pad})^{1/2}$ up to 601 GHz.

Together, these steps form a practical and physically grounded pathway toward flexible CNT RF transistors approaching, and potentially matching, the performance of their rigid counterparts.

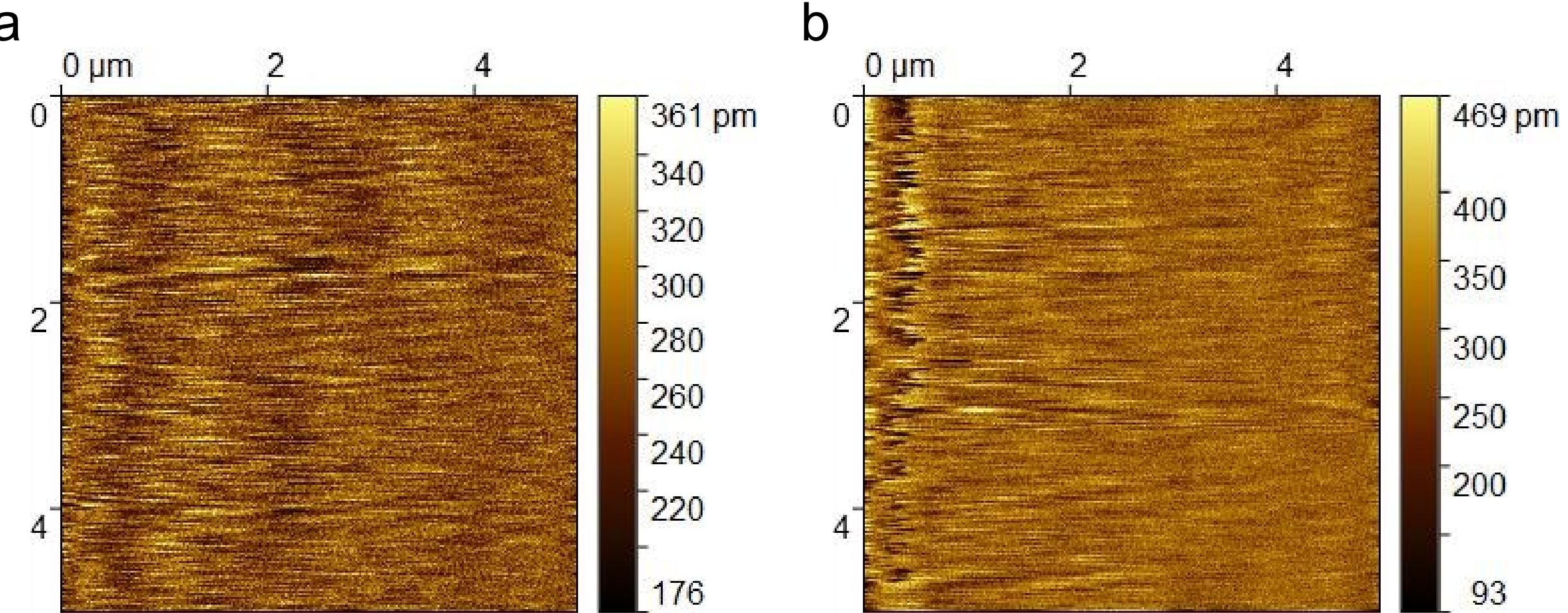

**Supplementary Fig. 1 | Atomic force microscopy (AFM) characterization of substrate surface roughness. a**, AFM image of the PI substrate surface over a 5 μm × 5 μm scan area, showing an average roughness of 0.27 nm. **b**, AFM image of a typical silicon substrate surface over a 5 μm × 5 μm scan area, showing an average roughness of 0.30 nm.

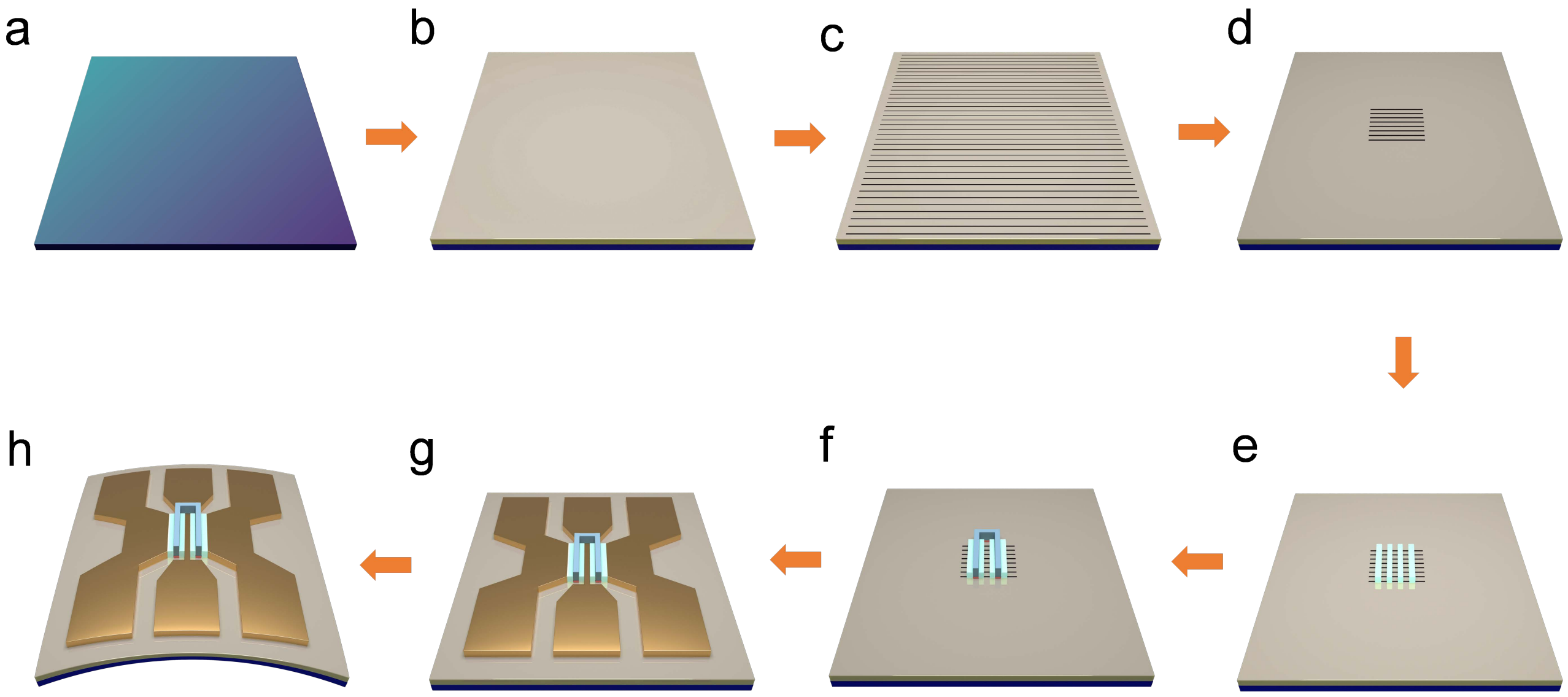

**Supplementary Fig. 2 | Fabrication process of flexible RF transistors. a**, Preparation of the rigid temporary supporting substrate (silicon wafer). **b**, Preparation of the flexible substrate (PI) with low dielectric loss and low surface roughness. **c**, Transfer of the aligned CNT arrays. **d**, Etching of the unwanted CNTs. **e**, Deposition of source and drain electrodes. **f**, Deposition of gate dielectrics and gate electrodes. **g**, Deposition of RF input and output pads. **h**, Exfoliation of the flexible electronic chip from the temporary supporting substrate.

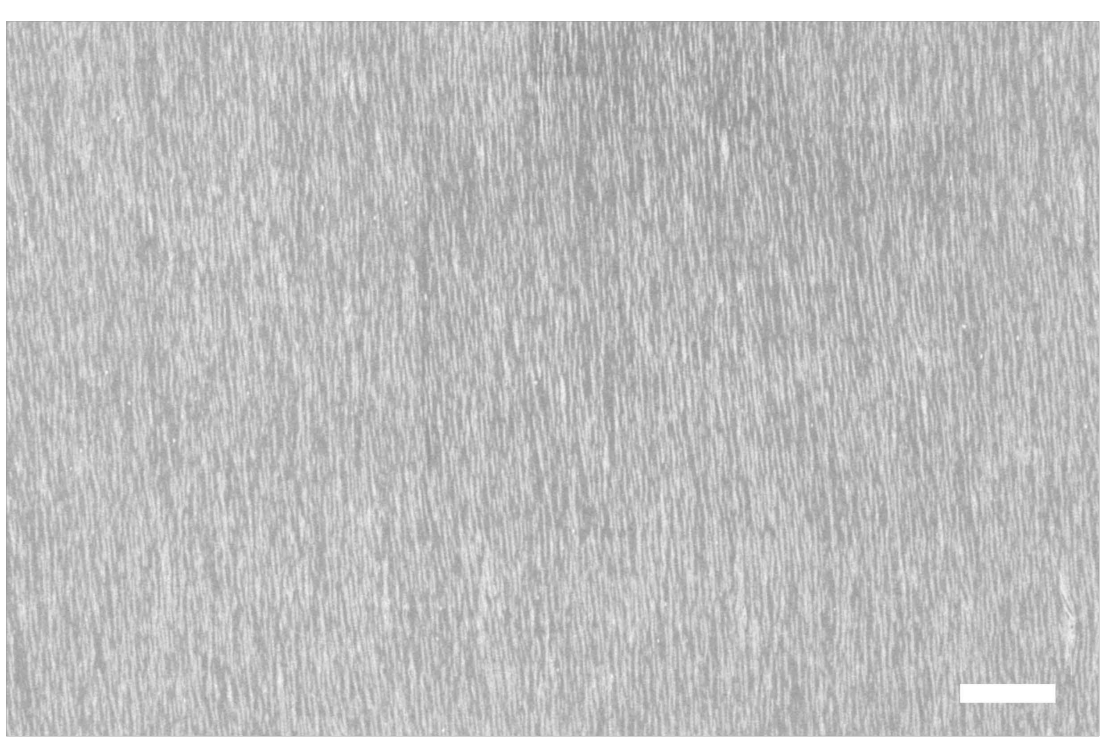

**Supplementary Fig. 3 | SEM image of the CNT material.** SEM image of aligned CNT arrays in the channel region of the flexible RF transistor. Scale bar: 1 μm.

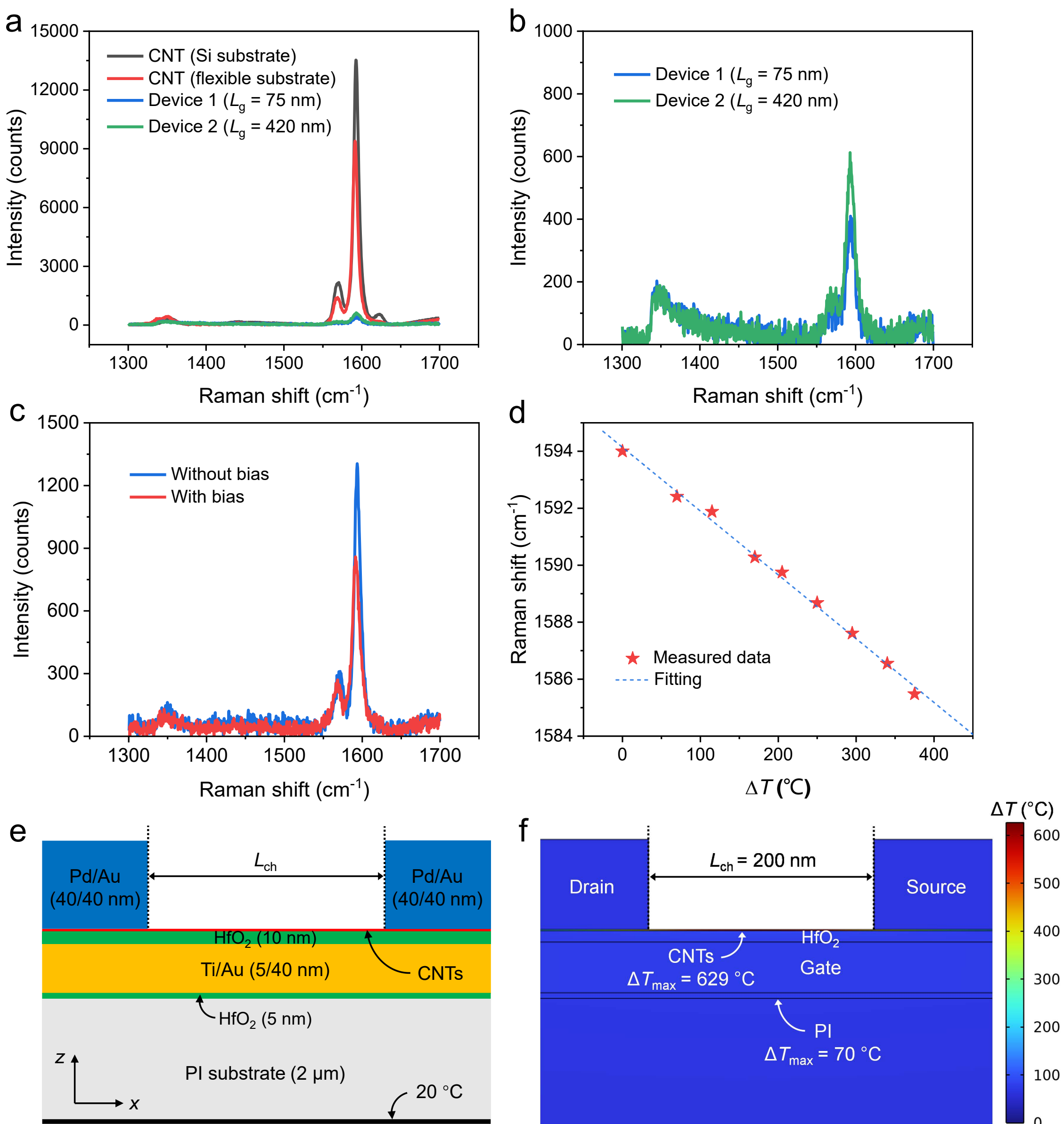

**Supplementary Fig. 4 | Raman thermometry and electro-thermal analysis of CNT films and flexible CNT transistors.** **a**, Raman spectra of CNTs on a silicon substrate, a flexible PI substrate, and within the channels of top-gated flexible CNT transistors with $L_g$ = 75 nm and 420 nm. Characteristic CNT peaks are observed at ~1350 $cm^{-1}$ (D band), ~1570 $cm^{-1}$ ($G^-$ band), and ~1592 $cm^{-1}$ ($G^+$ band). **b**, Rescaled Raman spectra of the CNTs in the channels of top-gated flexible CNT transistors shown in (**a**). **c**, Raman spectra acquired at the channel center of a bottom-gated flexible CNT transistor, measured with and without electrical bias ($I_d$ = 9.6 mA and 0 mA, correspondingly). The device geometry is shown in (**e**). **d**, Raman G-band shift of CNT films on a flexible substrate as a function of temperature, measured on a heating stage and fitted with a linear relationship to establish temperature calibration. **e**, Cross-sectional schematic of the bottom-gated flexible CNT transistor with $L_{ch}$ = 200 nm (not drawn to scale). The bottom surface of the PI substrate is fixed at room temperature (20 °C). **f**, Simulated temperature rise ($\Delta T$) of the bottom-gated flexible CNT transistor at a measured power dissipation of 7 mW, corresponding to the maximum sustainable power prior to electrical breakdown. All Raman spectra were acquired using a 532 nm excitation laser with an accumulation time of 20 s.

**a**

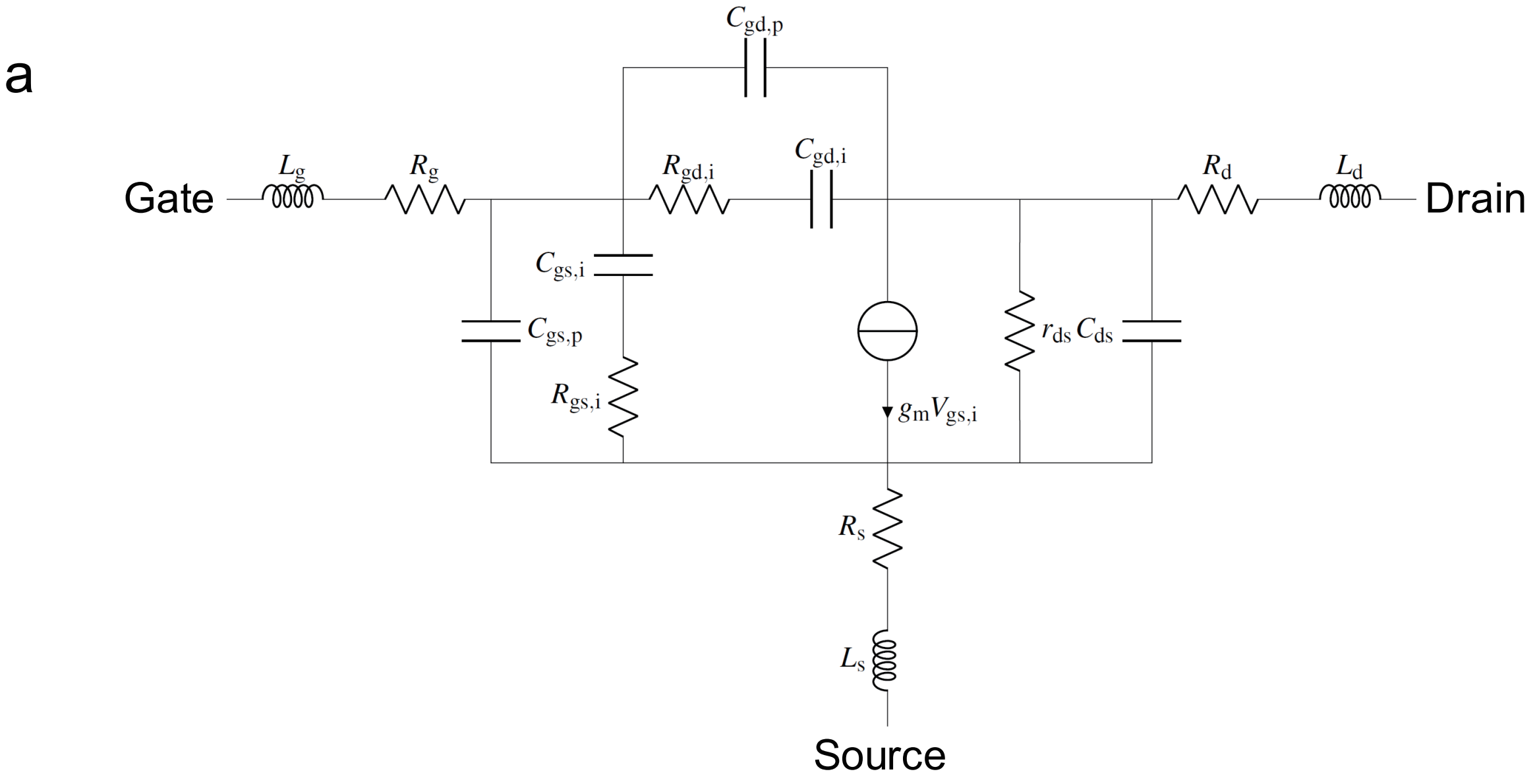

**b**

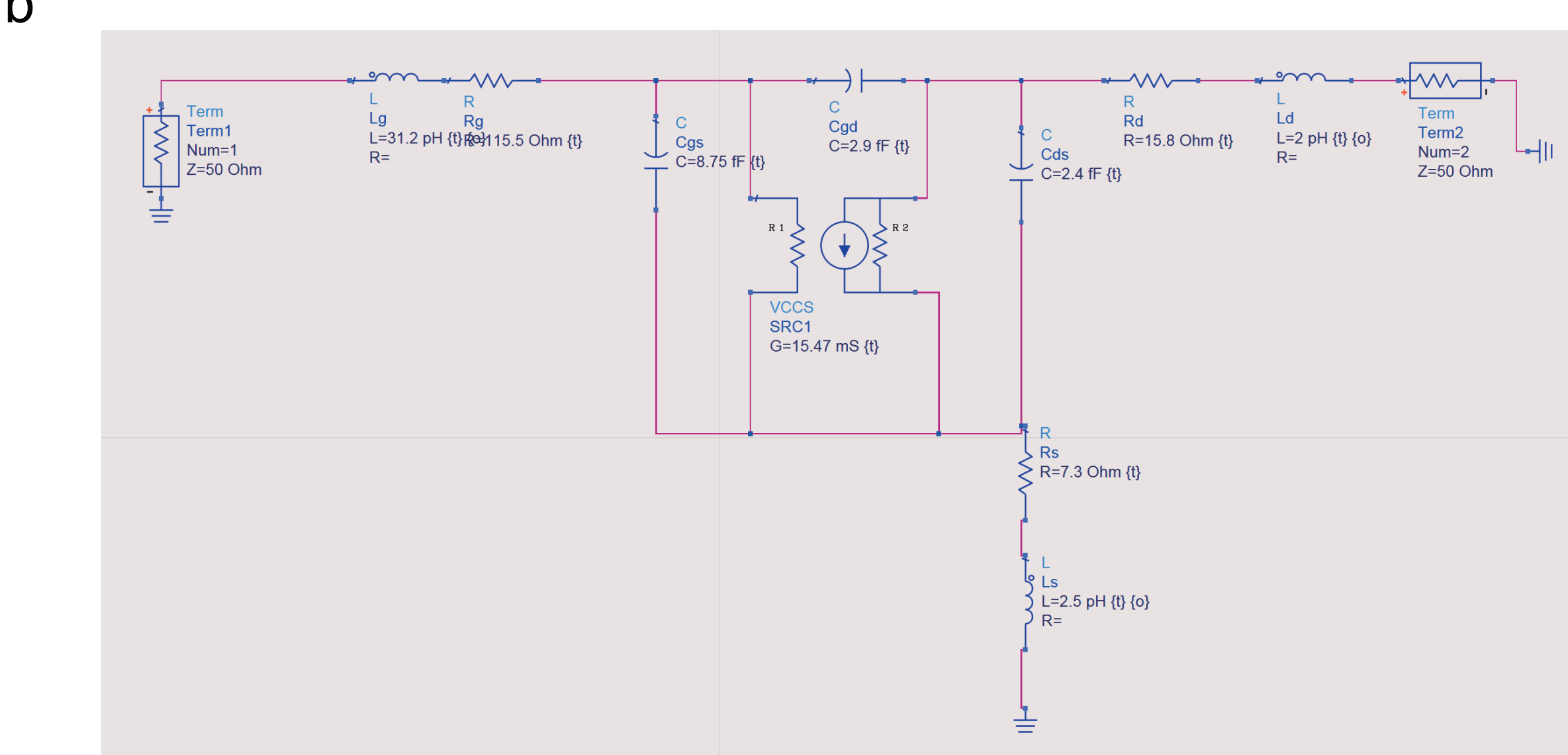

**Supplementary Fig. 5 | Small signal equivalent circuit modeling of flexible CNT-based RF transistors. a**, Circuit diagram of the small signal equivalent circuit model of pad de-embedded flexible CNT-based RF transistors. $L_d$, $L_s$, $L_g$, $R_d$, $R_s$, and $R_g$ represent drain serial inductance, source serial inductance, gate serial inductance, drain serial resistance, source serial resistance, and gate resistance, respectively. $C_{gs,i}$, $C_{gd,i}$, $C_{gs,p}$, $C_{gd,p}$, $C_{ds}$, $R_{gd,i}$, $R_{gs,i}$, $g_m$, $V_{gs,i}$, and $r_{ds}$ represent intrinsic gate-to-source capacitance, intrinsic gate-to-drain capacitance, parasitic gate-to-source capacitance, parasitic gate-to-drain capacitance, drain-to-source capacitance, intrinsic gate-to-drain resistance, intrinsic gate-to-source resistance, transconductance, the voltage on $C_{gs,i}$ divided from gate-to-source voltage, and drain-to-source output differential resistance, respectively. **b**, Schematic of small signal equivalent circuit model in Advanced Design System (ADS 2020, Keysight Technologies).

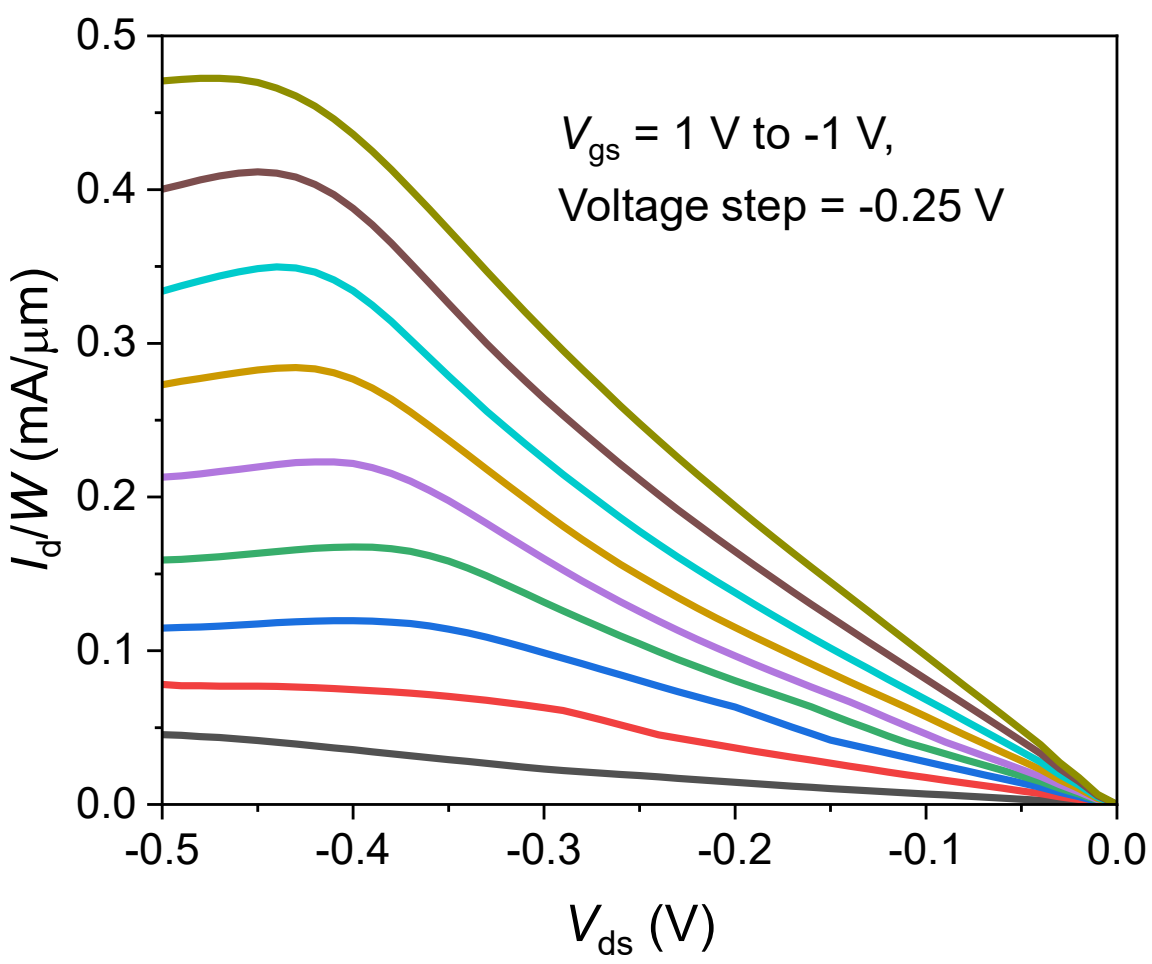

**Supplementary Fig. 6 | Output characteristics of a flexible CNT-based RF transistor ($L_g$ = 140 nm) fabricated without electro-thermal co-design.** Pronounced negative differential resistance due to severe self-heating at higher bias conditions significantly limits device performance and reliability.

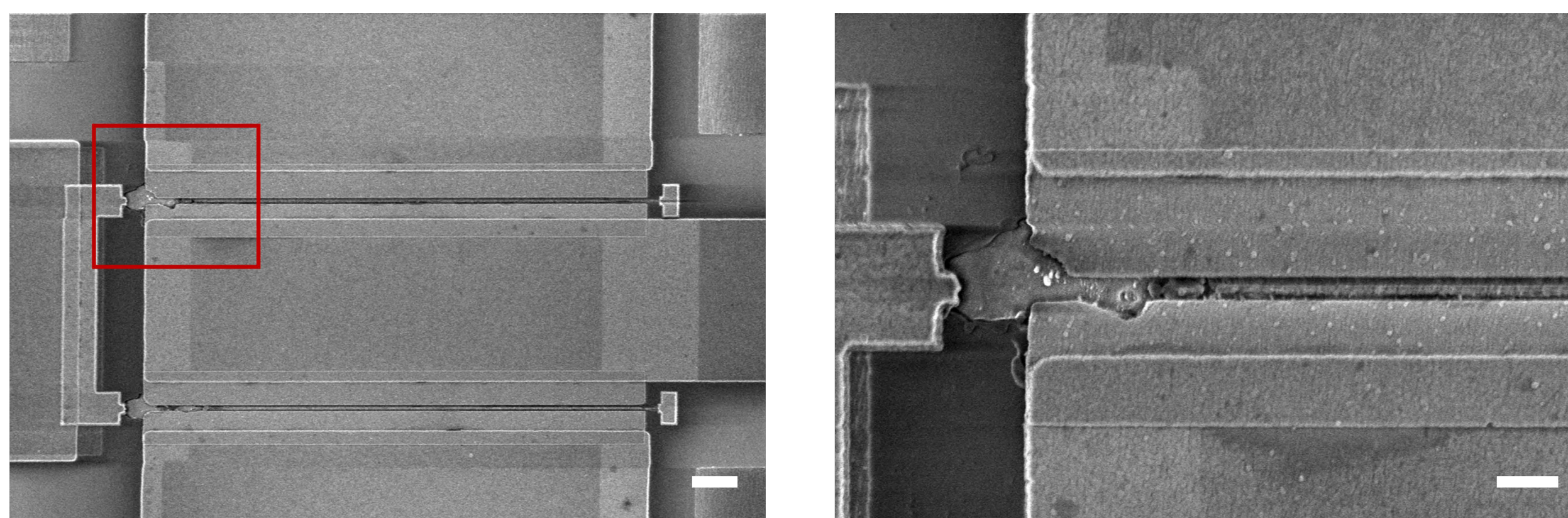

**Supplementary Fig. 7 | SEM images of a flexible CNT-based RF transistor fabricated without electro-thermal co-design after RF measurements. a**, SEM image of the damaged two-fingered flexible RF transistor. Scale bar: 1 μm. **b**, Zoomed-in SEM image of the area highlighted by the red box in (**a**). Scale bar: 400 nm.

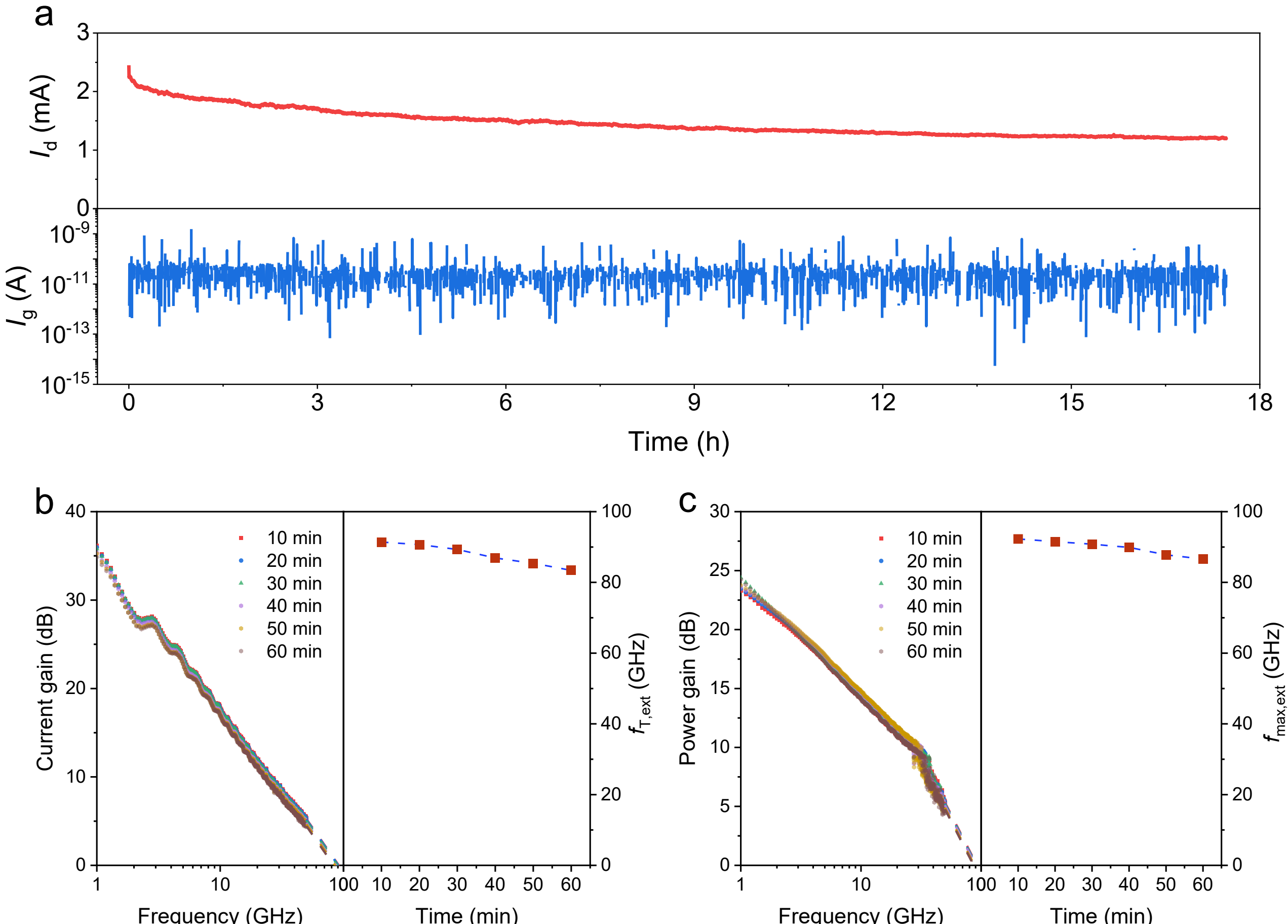

**Supplementary Fig. 8 | Stability test of flexible RF transistors with electro-thermal co-design. a,** Drain current (top) and gate current (bottom) of the same flexible RF transistor in Fig. 2e ($L_g$ = 160 nm) under constant bias conditions ($V_{gs}$ = -1 V and $V_{ds}$ = -0.7 V) as a function of time. The drain current gradually decreases from 2.24 mA to 1.90 mA after 1 h and to 1.21 mA after 17.5 h, while the gate current remains nearly unchanged. **b**, Current gain (left) and the corresponding $f_{T,ext}$ evolution over time (right) under continuous RF operation for the same flexible RF transistor ($L_g$ = 120 nm) shown in Fig. 3a,b. **c**, Power gain (left) and the corresponding $f_{max,ext}$ evolution over time (right) under continuous RF operation for the same device as in (**a**). The $f_{T,ext}$ and $f_{max,ext}$ only exhibited slightly decreases of 8.7% and 6.2%, respectively.

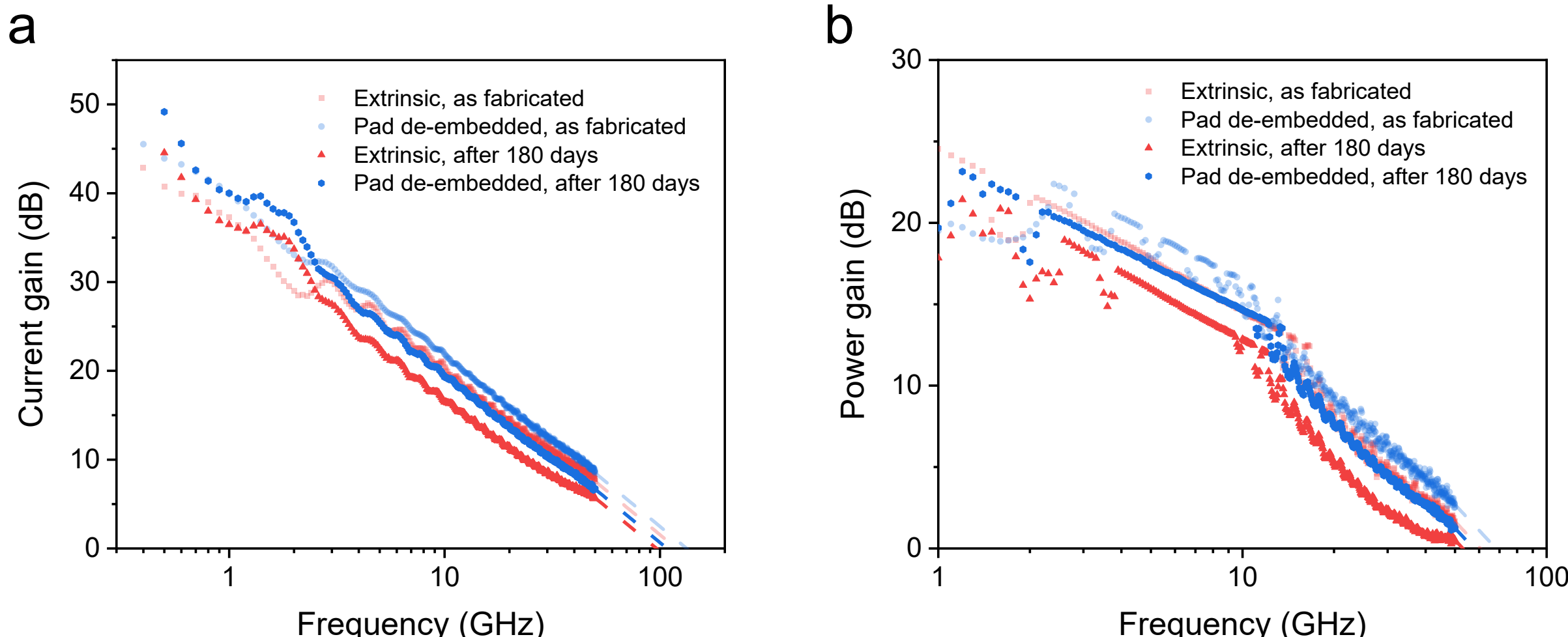

**Supplementary Fig. 9 | Shelf-life stability of flexible RF transistors. a**,**b**, Extrinsic and pad de-embedded current gain (**a**) and power gain (**b**) as a function of frequency for a representative flexible RF transistor with $L_g$ = 75 nm, measured immediately after fabrication and again after 180 days of storage in vacuum. Over this period, $f_{T,ext}$ and $f_{max,ext}$ decreased 19.3% (from 119.46 to 96.39 GHz) and 10.2% (from 58.78 to 53.04 GHz), respectively, while $f_{T,pad}$ and $f_{max,pad}$ decreased 18.8% (from 132.95 to 107.87 GHz) and 13.4% (from 66.89 to 57.67 GHz), respectively. These results demonstrate reasonably good shelf-life stability of our flexible RF transistors.

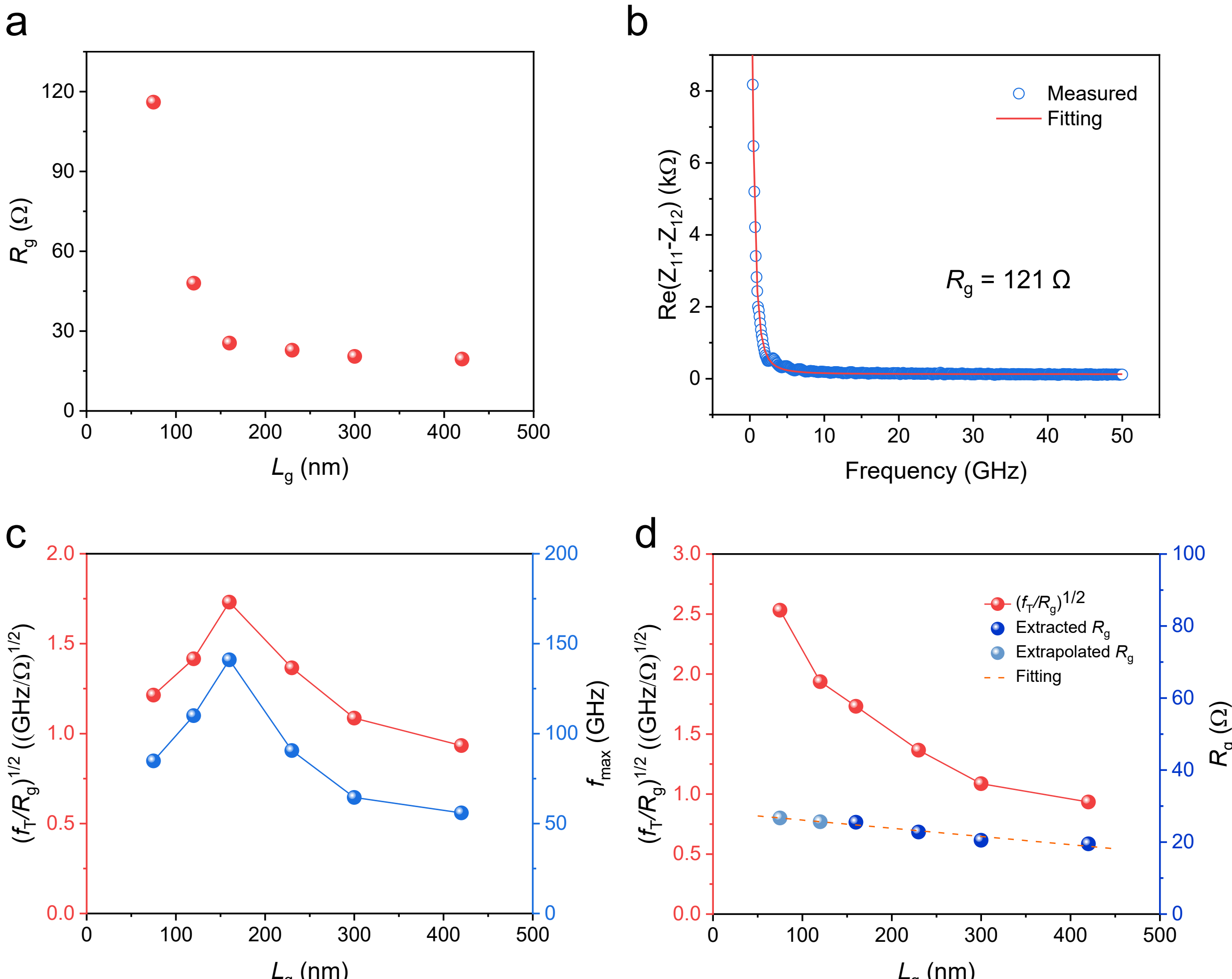

**Supplementary Fig. 10 | Extraction of $R_g$ in flexible RF transistors with different $L_g$ and its impact on $f_{max}$. a**, $R_g$ of flexible RF transistors with varying $L_g$, extracted using the small-signal equivalent circuit model in ADS. **b**, Extracted $R_g$ from the Z-parameters formulations based on the small-signal equivalent circuit model for a flexible RF transistor with $L_g = 75$ nm. **c**, $(f_T/R_g)^{1/2}$ as a function of $L_g$ for flexible RF transistors. The peak position matches the experimentally observed $f_{max}$ trend. **d**, Calculated $(f_T/R_g)^{1/2}$ assuming a gradual variation of $R_g$, obtained by linearly extrapolating $R_g$ from devices with $L_g \geq 160$ nm. Under this condition, $(f_T/R_g)^{1/2}$ increases monotonically with channel-length scaling.

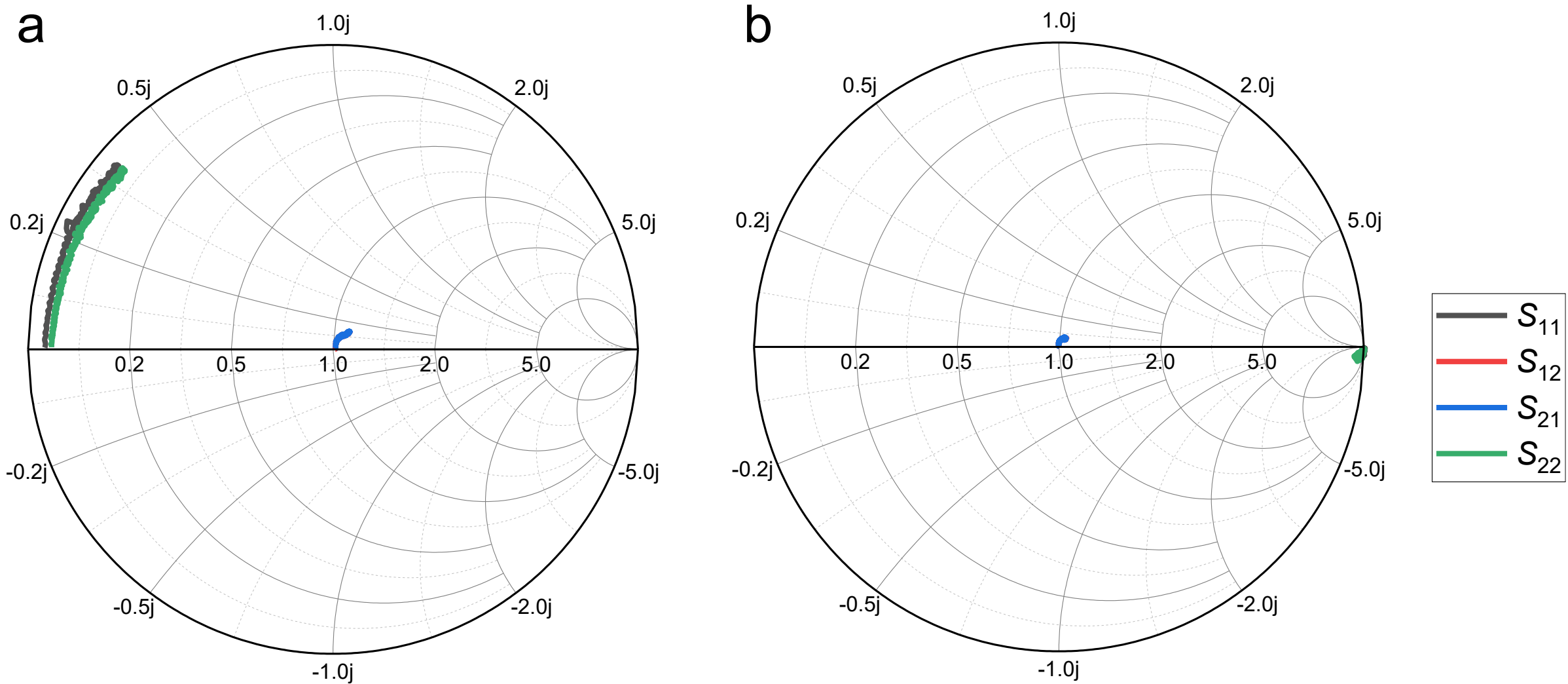

**Supplementary Fig. 11 | S-parameters of pad de-embedding structures. a**,**b**, S-parameters of the pad de-embedding short (**a**) and open structure (**b**) plotted in Smith charts.

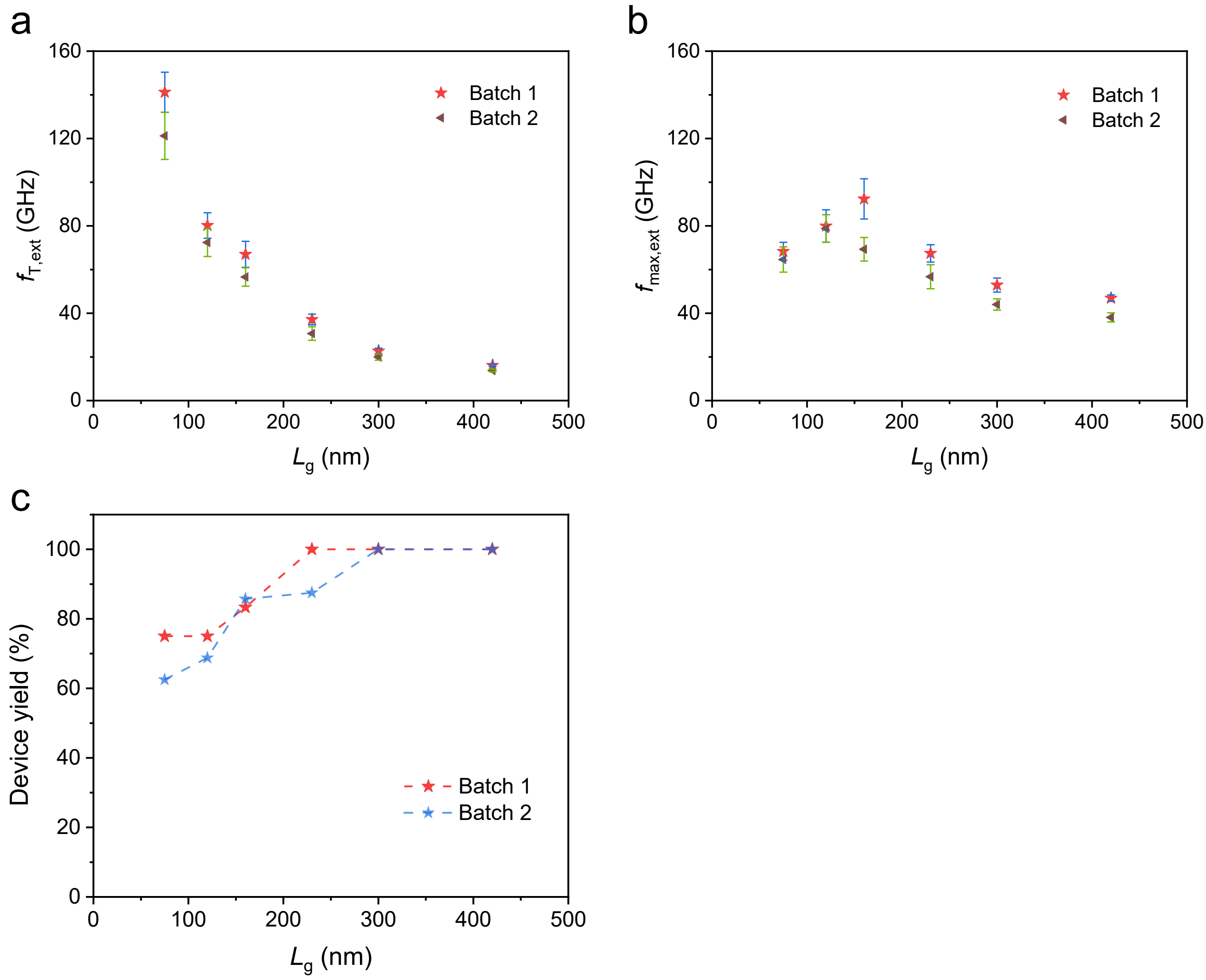

**Supplementary Fig. 12 | Process reliability and device yield of the flexible RF transistors. a**,**b**, Average $f_{T,ext}$ (**a**) and $f_{max,ext}$ (**b**) of multiple flexible RF transistors with varying $L_g$ from two different batches (72 devices in each batch), with error bars indicating standard deviations. **c**, Device yield of flexible RF transistors with varying $L_g$ from two different batches.

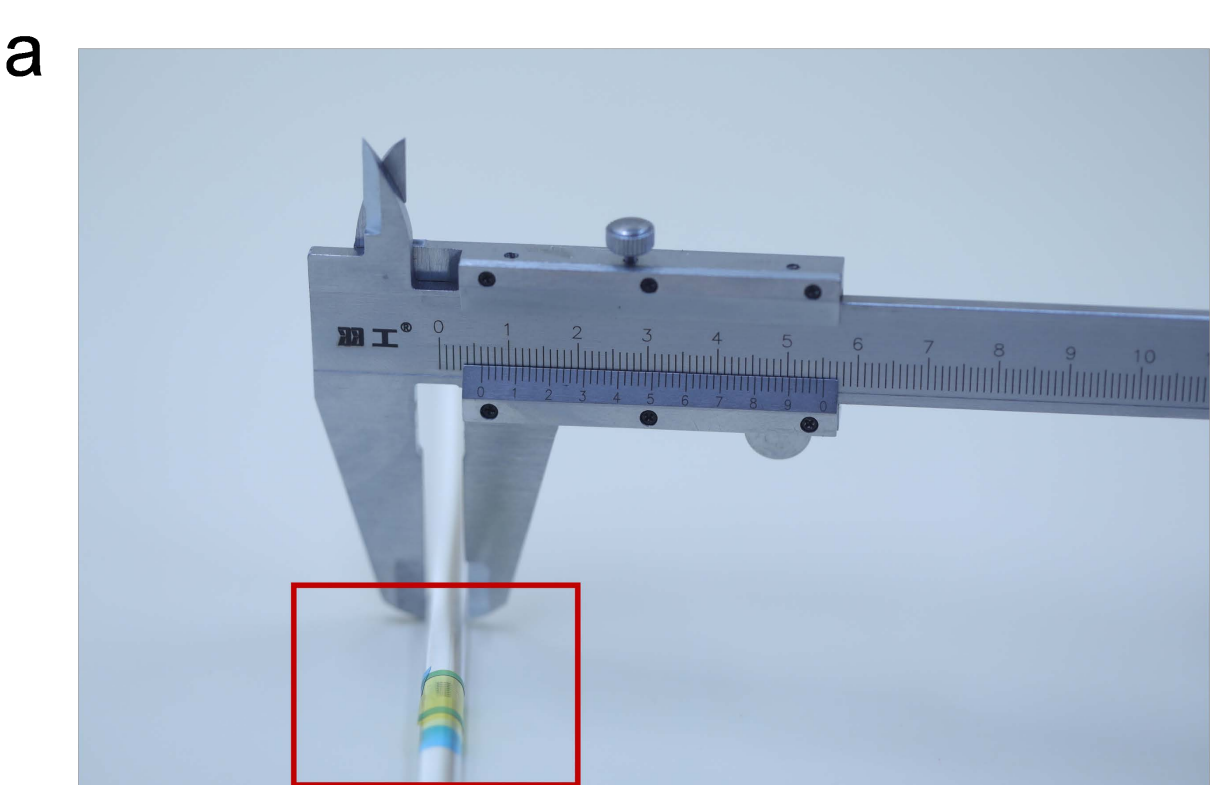

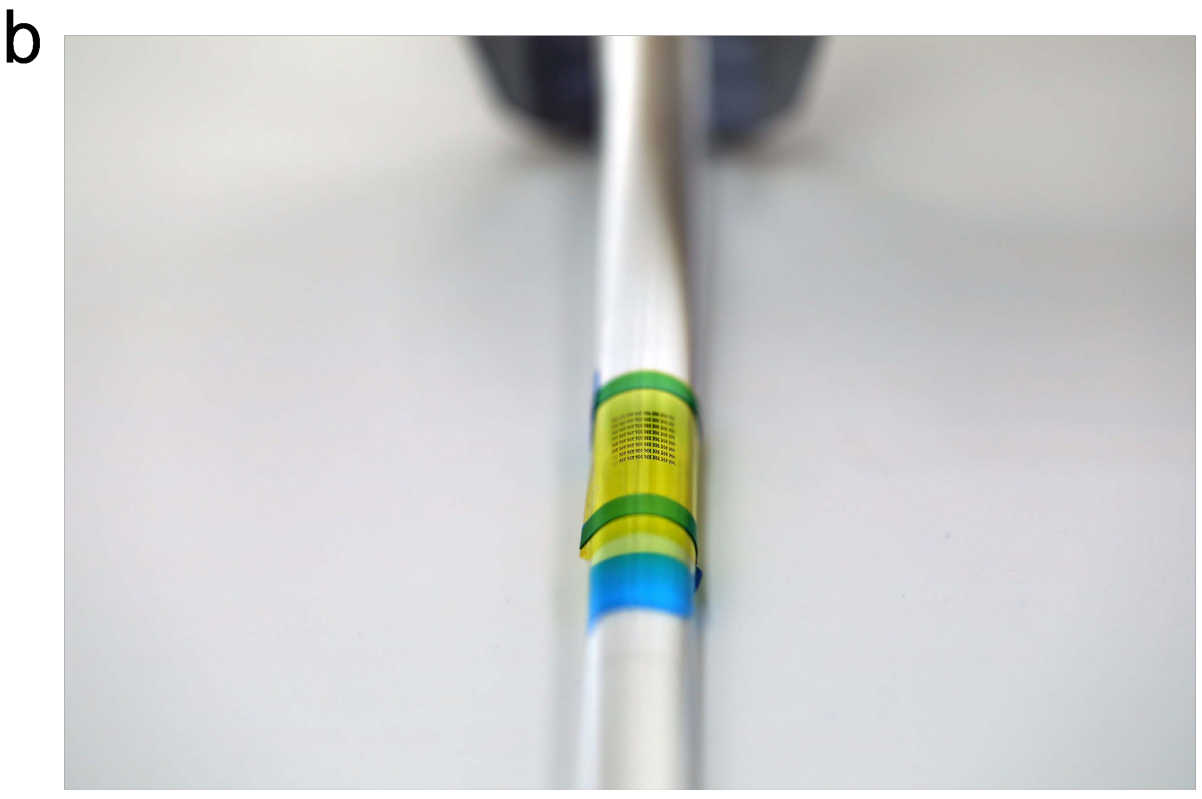

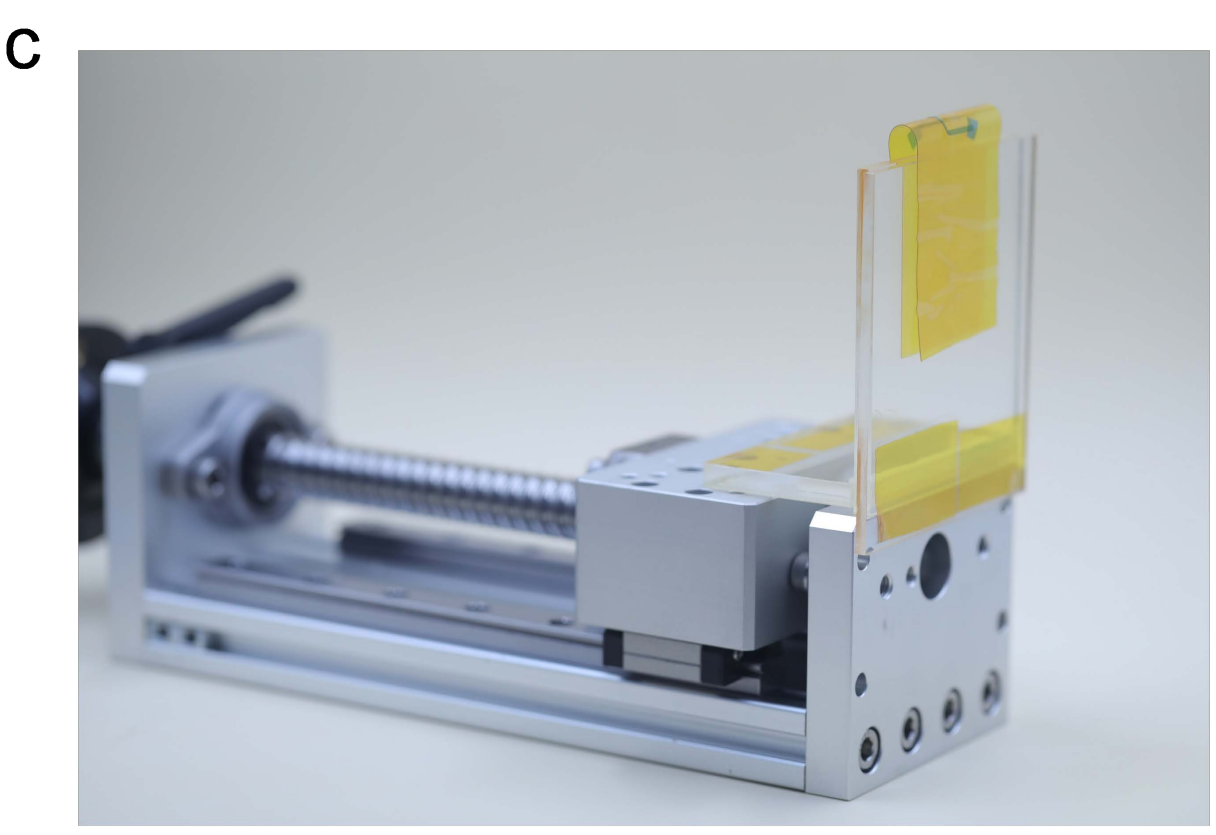

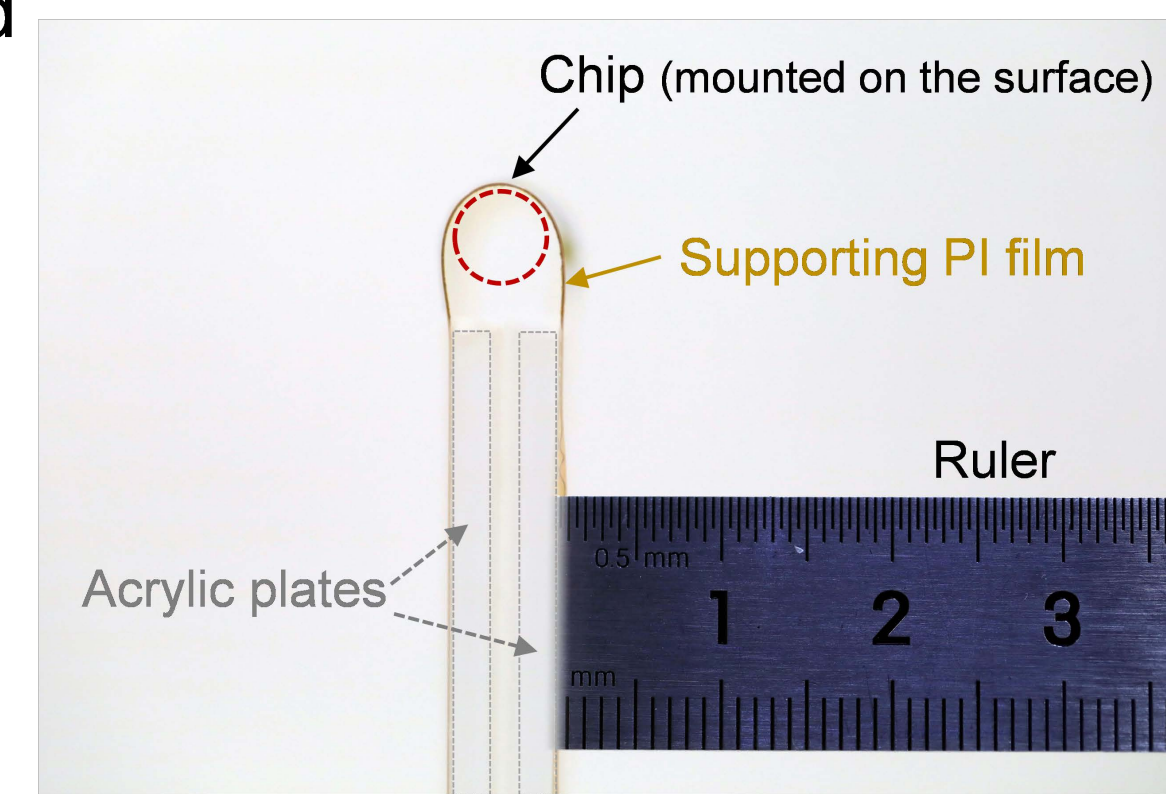

**Supplementary Fig. 13 | Testing setup for the flexibility assessment of the flexible RF transistors.** **a**, Photograph of the flexible electronic chip bent by attachment to an acrylic cylinder with a radius of 3 mm, measured using a micrometer. **b**, Enlarged view of the region highlighted by the red box in (**a**). **c**, Photograph of the cyclic bending test setup. **d**, Photograph of the cross-sectional view of the film under bending during the cyclic bending test.

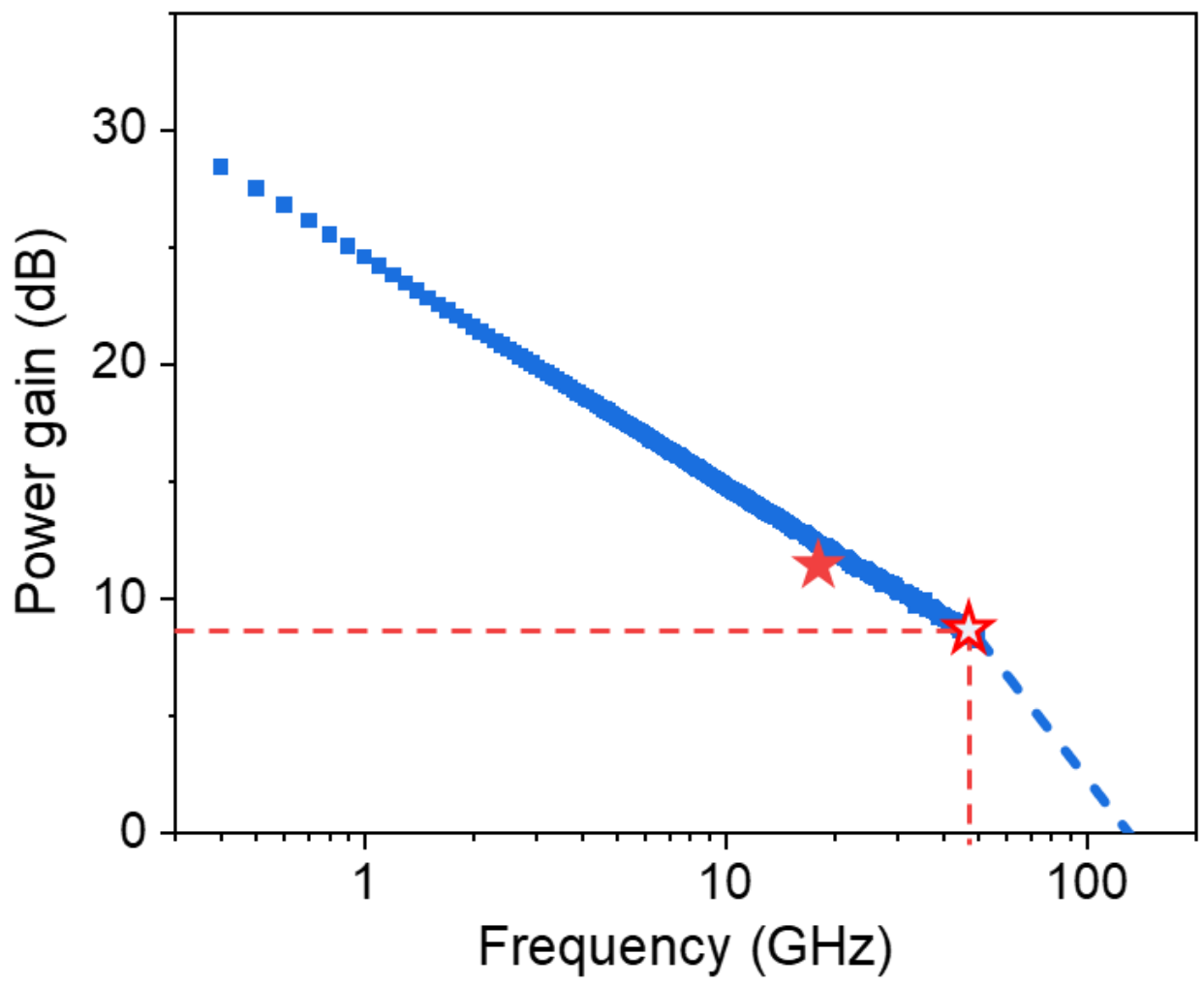

**Supplementary Fig. 14 | Pad de-embedded small-signal power gain as a function of frequency for the flexible power amplifier with $L_g$ = 160 nm.** The solid red star indicates the measured power gain of the amplifier at 18 GHz. The slope of the extrapolation line is -20 dB dec$^{-1}$. The extracted $f_{max,pad}$ is 131 GHz, and one-third of this value (indicated by the hollow red star) corresponds to a potential operating frequency of 43.6 GHz, at which the power gain is 8.9 dB.

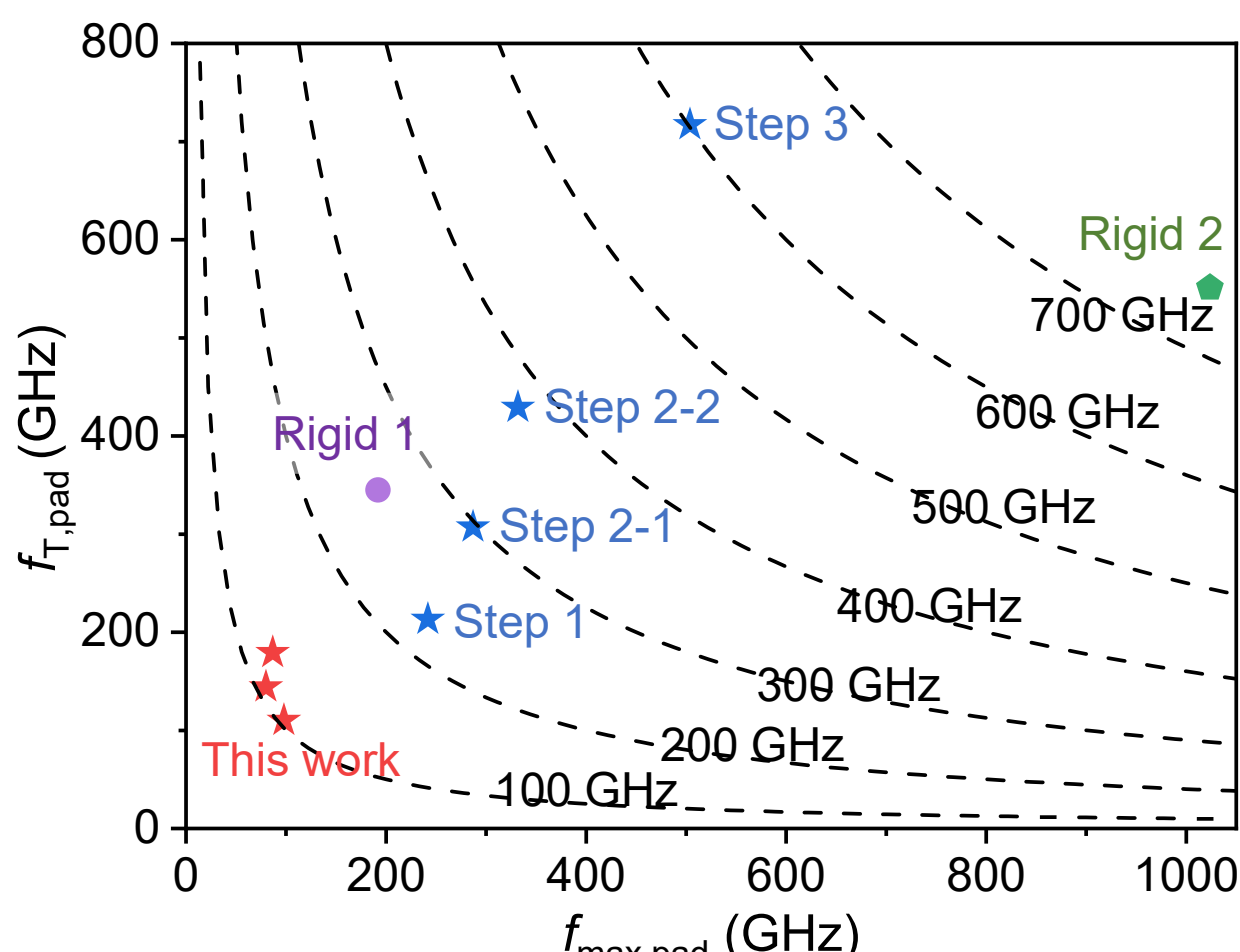

**Supplementary Fig. 15 | Roadmap illustrating the potential performance improvements of flexible RF transistors based on aligned CNT arrays, derived from ADS simulations following the stepwise electro–thermal–mechanical co-design strategy described in the text.** The simulations sequentially incorporate (Step 0) improved substrate properties, (Step 1) gate-structure optimization to reduce $R_g$, (Step 2) channel-length scaling (step 2-1: 50 nm; step 2-2: 30 nm) under controlled thermal/mechanical conditions, and (Step 3) enhancement of CNT array quality and density. All simulations are performed for devices with a channel width of 20 μm, and the corresponding key parameters are summarized in Supplementary Table 10.

**Supplementary Table 1 | Benchmarking of the DC performance of CNT-based MOSFETs and flexible MOSFETs based on various materials.**

| Materials | Type | $L_g$ (nm) | $\lvert V_{ds} \rvert$ (V) | $I_{on}$ (μA/μm) | $g_m$ (μS/μm) | Reference |
|---|---|---|---|---|---|---|
| CNT | flexible | 75 | 0.6 | 947 | 728 | This work |
| CNT | flexible | 450 | 1 | 188 | 123 | [34] |
| CNT | flexible | 2000 | 3 | 17 | 1.46 | [85] |
| CNT | flexible | 5000 | 2 | 6.05 | 5.45 | [86] |
| CNT | flexible | 3000 | 5 | 15 | 4 | [26] |
| CNT | rigid | 90 | 0.9 | 1920 | 1400 | [27] |
| CNT | rigid | 100 | 1.5 | 350 | 310 | [53] |
| CNT | rigid | 110 | 1 | 270 | 215 | [52] |
| CNT | rigid | 200 | 0.8 | 1180 | 1000 | [87] |
| CNT | rigid | 100 | 1 | 1300 | 900 | [32] |
| Graphene | flexible | 100 | 0.5 | 850 | 250 | [16] |
| Graphene | flexible | 300 | 0.3 | 625 | 423 | [88] |
| Graphene | flexible | 50 | 0.5 | 570 | 65 | [15] |
| Graphene | flexible | 200 | 0.75 | 630 | 656 | [89] |
| $MoS_2$ | flexible | 82 | 1 | 466 | 24 | [37] |
| $MoS_2$ | flexible | 116 | 2 | 48 | 18 | [90] |
| $MoS_2$ | flexible | 6000 | 5 | 66 | 55 | [91] |
| BP | flexible | 500 | 2 | 100 | 41.3 | [17] |
| IGZO | flexible | 160 | 1 | 250 | 7.4 | [38] |
| IGZO | flexible | 500 | 5 | 2 | 1.48 | [13] |
| ITO | flexible | 15 | 0.5 | 428 | 580 | [14] |
| ITO | flexible | 160 | 1.5 | 83.2 | 153 | [92] |
| Organic | flexible | 600 | 3 | 5 | 6.4 | [12] |
| Organic | flexible | 1000 | 0.5 | 56 | 70.5 | [39] |

**Supplementary Table 2 | Electro-thermal design parameters and their effects on thermal behavior and RF performance.** Electro-thermal simulations were conducted using Advanced Design System (ADS) and COMSOL Multiphysics for a representative flexible RF transistor with $L_g$ = 75 nm ($L_{ch}$ = 120 nm). Key design parameters include the thickness of the source, drain, and gate electrodes, as well as the $L_g/L_{ch}$ ratio, and their corresponding impacts on maximum temperature rise in the channel and RF performance metrics.

| Line | Design parameters (S\|D\|G thickness \| $L_g/L_{ch}$ ratio) | $f_{T,pad}$ (GHz) | $f_{max,pad}$ (GHz) | $\Delta T_{max,CNT}$ (°C) |
|---|---|---|---|---|
| 1 | 20\|20\|190 nm \| 0.625 | 179 | 72 | 331 |
| 2 | 10\|10\|190 nm \| 0.625 | 184 | 74.8 | 398 |
| 3 | 40\|40\|190 nm \| 0.625 | 169 | 68.4 | 298 |
| 4 | 60\|60\|190 nm \| 0.625 | 161 | 64.6 | 288 |
| 5 | 80\|80\|190 nm \| 0.625 | 154 | 61.2 | 283 |
| 6 | 100\|100\|190 nm \| 0.625 | 146 | 58 | 281 |
| 7 | 20\|20\|400 nm \| 0.625 | 179 | 107 | 322 |
| 8 | 20\|20\|250 nm \| 0.625 | 179 | 78 | 328 |
| 9 | 20\|20\|100 nm \| 0.625 | 179 | 55 | 343 |
| 10 | 20\|20\|40 nm \| 0.625 | 179 | 33 | 354 |
| 11 | 20\|20\|190 nm \| 0.417 | 173 | 72 | 337 |
| 12 | 20\|20\|190 nm \| 0.750 | 158 | 66 | 330 |
| 13 | 20\|20\|190 nm \| 0.917 | 125 | 52 | 328 |

**Supplementary Table 3 | Comparison of the extrapolated $f_{T,pad}$, $f_{max,pad}$ and the simulated $f_{T,pad}$, $f_{max,pad}$ using small signal equivalent model for devices with different $L_g$.** The two methods show excellent agreement.

| $L_g$ (nm) | Extrapolated $f_{T,pad}$ (GHz) | Simulated $f_{T,pad}$ (GHz) | Extrapolated $f_{max,pad}$ (GHz) | Simulated $f_{max,pad}$ (GHz) |
|---|---|---|---|---|
| 75 | 171 | 179 | 84.8 | 86.5 |
| 120 | 96.2 | 98 | 110 | 108 |
| 160 | 76.4 | 80.8 | 141 | 143.7 |
| 230 | 42.5 | \ | 90.6 | 90 |
| 300 | 24.2 | \ | 64.5 | 64.1 |
| 420 | 17 | \ | 55.9 | 55.8 |

**Supplementary Table 4 | Comparison of the extrapolated $f_{T,ext}$ and the calculated $f_{T,ext}$ using Gummel's method[38,90,93] for devices with different $L_g$.** The two methods show excellent agreement.

| $L_g$ (nm) | 75 | 120 | 160 | 230 | 300 | 420 |
|---|---|---|---|---|---|---|
| Extrapolated (GHz) | 152 | 91.4 | 73.8 | 40.6 | 23.7 | 16.5 |
| Calculated (GHz) | 153 | 91.7 | 74.0 | 40.4 | 23.6 | 16.7 |

**Supplementary Table 5 | Benchmarking RF performance for CNT-based MOSFETs fabricated on flexible and rigid substrates, as well as flexible MOSFETs based on other various materials.**

| Materials | Type | $L_g$ (nm) | $|V_{ds}|$ (V) | $f_{T,ext}$ (GHz) | $f_{max,ext}$ (GHz) | $f_{T,pad}/f_{T,ext}$ | $f_{max,pad}/f_{max,ext}$ | Reference |
|---|---|---|---|---|---|---|---|---|
| CNT | flexible | 75 | 0.65 | 152 | 73.5 | 1.12 | 1.15 | This work |
| CNT | flexible | 120 | 0.95 | 91.4 | 92.3 | 1.05 | 1.19 | This work |
| CNT | flexible | 160 | 1.05 | 73.8 | 102 | 1.04 | 1.38 | This work |
| CNT | flexible | 1000 | 3 | 1.6 | 1.8 | 1.2 | 1.2 | [29] |
| CNT | flexible | 800 | 0.7 | 1 | \ | 5 | \ | [94] |
| CNT | rigid | 100 | 1.5 | 40 | 40 | 2.0 | 1.8 | [53] |
| CNT | rigid | 110 | 1.5 | 75 | 102 | \ | \ | [52] |
| CNT | rigid | 50 | 1.4 | 186 | 158 | \ | \ | [27] |
| CNT | rigid | 4000 | \ | 2.5 | 1.1 | \ | \ | [25] |
| CNT | rigid | 120 | 1.8 | 12 | 28 | 2.5 | 1.5 | [95] |
| CNT | rigid | 300 | 1.5 | 4 | \ | \ | \ | [24] |
| Graphene | flexible | 100 | 0.5 | 39 | 13.5 | \ | \ | [16] |
| Graphene | flexible | 300 | 0.3 | 31 | 28 | \ | \ | [88] |
| Graphene | flexible | 50 | \ | 56 | 14.5 | \ | \ | [15] |
| Graphene | flexible | 260 | 0.5 | 23.6 | 6.5 | 1.64 | 1.2 | [96] |
| Graphene | flexible | 200 | 0.7 | 32 | 20 | 2.0 | 1.7 | [89] |
| $MoS_2$ | flexible | 300 | 4 | 4 | 9 | \ | \ | [18] |
| $MoS_2$ | flexible | 68 | 8 | 4.7 | 5.4 | 2.9 | 1.9 | [90] |
| BP | flexible | 500 | 1.8 | 7 | 10.3 | 2.1 | 1.7 | [17] |
| IGZO | flexible | 500 | 2~5 | 0.135 | 0.304 | \ | \ | [13] |
| IGZO | flexible | 160 | 0.5 | 0.0059 | 0.0192 | \ | \ | [38] |
| ITO | flexible | 160 | 1 | 2.1 | 3.7 | \ | \ | [92] |
| ITO | flexible | 15 | 0.5 | 8.2 | 14.4 | 1.44 | 1.04 | [14] |
| Organic | flexible | 600 | 3 | 0.021 | \ | \ | \ | [12] |
| Organic | flexible | 1200 | 15 | 0.019 | \ | \ | \ | [97] |

**Supplementary Table 6 | Average values and standard deviations of the extrinsic and pad de-embedded $f_T$ and $f_{max}$.**

| $L_g$ (nm) | $f_{T,ext}$ (GHz) | $f_{max,ext}$ (GHz) | $f_{T,pad}$ (GHz) | $f_{max,pad}$ (GHz) |
|---|---|---|---|---|
| 75 | 141.17±9.20 | 68.33±4.13 | 158.67±7.92 | 78.50±4.59 |
| 120 | 80.17±5.84 | 79.93±7.44 | 83.67±6.34 | 97.36±11.46 |
| 160 | 66.96±5.95 | 92.34±9.23 | 69.05±6.29 | 127.24±17.09 |
| 230 | 37.11±2.48 | 67.42±3.98 | 39.14±2.47 | 81.58±5.83 |

**Supplementary Table 7 | Device characteristics used in the finite-element electro-thermal simulations.** All devices have a channel width $W$ of 10 μm.

| $L_g$ (nm) | $L_{ch}$ (nm) | $f_{T,ext}$ (GHz) | $\|V_{ds}\|$ (V) | $I_d$ (mA) | $P_{dc}$ (mW) |
|---|---|---|---|---|---|
| 75 | 120 | 152 | 0.65 | 3.07 | 1.99 |
| 120 | 235 | 91.4 | 0.95 | 2.38 | 2.26 |
| 160 | 265 | 73.8 | 1.05 | 2.24 | 2.35 |
| 230 | 340 | 40.6 | 1.15 | 1.62 | 1.86 |
| 300 | 450 | 23.7 | 1.5 | 1.30 | 1.96 |
| 420 | 540 | 16.5 | 1.8 | 1.23 | 2.22 |

**Supplementary Table 8 | Nominal thermal properties used in simulations.** Some thermal boundary conductance (TBC) values not available in the literature were approximated by TBCs for pairs of similar and/or better-studied materials, using the lower bound estimates to model the worst-case scenario. The actual thermal properties vary depending on the material and interface quality, and deposition conditions. The thermal conductivity (TC) of Au as source and drain is lowered due to its thinner film, on the other hand, the TC of the Au gate is close to the bulk value[45]. We also account for the anisotropic TC of the PI substrate.

| Material | TC ($Wm^{-1}K^{-1}$) | Material Interface | TBC ($MWm^{-2}K^{-1}$) |
| --- | --- | --- | --- |
| PI[98-100] | 0.2 (Cross-plane)<br>1.1 (In-plane) | CNT-$HfO_2$ and<br>CNT-$Al_2O_3$[44,49,68,101,102] | 50 |
| Au[45] | 300 (Gate)<br>150 (Source and Drain) | CNT-Pd/Au[44,103] | 50 |
| $HfO_2$[46] | 1.0 | $HfO_2$-PI[37,104] | 2.5 |
| $Al_2O_3$[46,47,105] | 1.5 | $HfO_2$-Au/Pd[41,106] | 50 |
| CNT[63] | 50 (In-plane)<br>$5.7 \times 10^{-4}$ (Cross-plane) | $Al_2O_3$-Al[47,107] | 200 |
| Al[45] | 60 (Gate) | | |

**Supplementary Table 9 | Comparison of $R_g$ extracted using ADS software ($R_{g,sim}$) and obtained from fitting the Z-parameter formulations ($R_{g,cal}$).**

| $L_g$ (nm) | 75 | 120 | 160 | 230 | 300 | 420 |
|---|---|---|---|---|---|---|
| $R_{g,sim}$ (GHz) | 116 | 46.8 | 25.5 | 22.8 | 20.5 | 19.5 |
| $R_{g,cal}$ (GHz) | 121 | 45.0 | 26.3 | 21.7 | 19.8 | 18.4 |

**Supplementary Table 10 | Key parameters used in the ADS simulations to evaluate the potential performance improvements of flexible RF transistors based on aligned CNT arrays.** All our simulated devices have a channel width $W$ = 20 μm. The Rigid 1 device[27] has $W$ = 20 μm, while the Rigid 2 device[84] has $W$ = 100 μm.

| Step | $L_g$ (nm) | $R_g$ (Ω) | $C_{gs}$ (fF) | $C_{gd}$ (fF) | $r_{ds}$ (Ω) | $g_m$ (mS) | $f_{T,pad}$ (GHz) | $f_{max,pad}$ (GHz) |
|---|---|---|---|---|---|---|---|---|
| Rigid 1[27] | 50 | 21.1 | 7.61 | 1.88 | 57.6 | 24.8 | 345 | 206 |
| Rigid 2[84] | 35 | 3.4 | 9.64 | 2.57 | 45 | 158 | 551 | 1024 |
| 0 | 75 | 116 | 8.75 | 2.90 | 104 | 15.5 | 179 | 86.8 |
| 1 | 75 | 10 | 8.75 | 2.90 | 104 | 15.5 | 213 | 242 |
| 2-1 | 50 | 10 | 7.16 | 2.50 | 75 | 18 | 307 | 287 |
| 2-2 | 30 | 10 | 5.77 | 2.39 | 60 | 20 | 429 | 332 |
| 3 | 30 | 10 | 5.77 | 2.39 | 40 | 30 | 717 | 504 |

**Supplementary References**